\documentclass[letterpaper]{aa}
\usepackage{graphicx}
\usepackage{txfonts}
\begin{document}
\title{
Temporal variations of the outer atmosphere and the dust shell 
of the carbon-rich Mira variable V~Oph probed with VLTI/MIDI
\thanks{Based on observations made with the Very Large Telescope 
Interferometer of the European Southern Observatory. 
Program ID: 075.D-0607}
\fnmsep
\thanks{Table.~4 is only available in electronic form at the CDS via
anonymous ftp to cdsarc.u-strasbg.fr (130.79.128.5) or via 
http://cdsweb.u-strasbg.fr/cgi-bin/qcat?J/A+A}
}

\author{K.~Ohnaka\inst{1} 
\and
T.~Driebe\inst{1} 
\and
G.~Weigelt\inst{1} 
\and
M.~Wittkowski\inst{2}
}

\offprints{K.~Ohnaka}

\institute{
Max-Planck-Institut f\"{u}r Radioastronomie, 
Auf dem H\"{u}gel 69, D-53121 Bonn, Germany\\
\email{kohnaka@mpifr-bonn.mpg.de}
\and
European Southern Observatory, Karl-Schwarzschild-Str.~2, 
D-85748 Garching, Germany
}

\date{Received / Accepted }

\abstract
{}
{We present the first multi-epoch $N$-band spectro-interferometric 
  observations of the 
  carbon-rich Mira variable V~Oph using MIDI at the ESO's 
  Very Large Telescope Interferometer.  Our aim is to study temporal 
  variations of physical properties of the outer atmosphere and 
  the circumstellar dust shell based on spectrally-dispersed 
  $N$-band visibilities over the \mbox{C$_2$H$_2$}\ (+HCN) features and the 
  dust emission.  
}
{Our MIDI observations were carried out at three different 
  phases 0.18, 0.49, and 0.65, 
  with three different baselines (projected baseline lengths of 
  42--124~m) using four 8.2~m Unit Telescopes (UT2-UT4, UT1-UT4, 
  and UT2-UT3 baseline configurations).  
}
{The wavelength dependence of the uniform-disk diameters 
  obtained at all epochs 
  is characterized by a roughly constant region between 8
  and 10~\mbox{$\mu$m}\ with a slight dip centered at $\sim$9.5~\mbox{$\mu$m}\ 
  and a gradual increase longward of 10~\mbox{$\mu$m}.  These $N$-band 
  angular sizes are significantly larger than the estimated 
  photospheric size of V~Oph.  
  The angular sizes observed at different epochs reveal that 
  the object appears smaller at phase 0.49 (minimum light) with 
  uniform-disk diameters of $\sim$5--12~mas than at phases 0.18 
  ($\sim$12--20~mas) and 0.65 ($\sim$9--15~mas). 
  We interpret these results with a model consisting of 
  optically thick \mbox{C$_2$H$_2$}\ layers and an optically thin dust shell.  
  Our modeling suggests that the \mbox{C$_2$H$_2$}\ layers around V~Oph 
  are more extended ($\sim$1.7--1.8~\mbox{$R_{\star}$}) at phases 0.18 and 0.65 
  than at phase 0.49 ($\sim$1.4~\mbox{$R_{\star}$}) and that 
  the \mbox{C$_2$H$_2$}\ column densities appear to be the smallest at phase 
  0.49. 
  We also find that the dust shell consists of amorphous 
  carbon and SiC with an inner radius of 
  $\sim$2.5~\mbox{$R_{\star}$}, and the total optical depths of 
  $\mbox{$\tau_{V}$} \! \approx \! 0.6$--0.9 ($\tau_{11.3\mu{\rm m}} \! \approx \!
  0.003$ and 0.004 for amorphous carbon and SiC, respectively) found 
  at phases 0.18 and 0.65 
  are higher than the value obtained at phase 0.49, 
  $\mbox{$\tau_{V}$} \! \approx \! 0.3$ ($\tau_{11.3\mu{\rm m}} \! \approx \! 
  0.001$ and 0.002 for amorphous carbon and SiC, respectively).
}
{ 
  Our MIDI observations and modeling indicate that carbon-rich Miras 
  also have extended layers of polyatomic molecules as previously 
  confirmed in oxygen-rich Miras.  
  The temporal variation of the 
  $N$-band angular size is largely governed by the variations of 
  the opacity and the geometrical extension of the \mbox{C$_2$H$_2$}\ layers 
  and the dust shell, and consequently, this masks the size variation
  of the photosphere.  Also, the observed weakness of the 
  mid-infrared \mbox{C$_2$H$_2$}\ absorption in carbon-rich Miras can be explained 
  by the emission from the extended \mbox{C$_2$H$_2$}\ layers and the dust shell.  
}


\titlerunning{VLTI/MIDI observations of the carbon-rich Mira V~Oph}
\authorrunning{Ohnaka et al.}
\maketitle

\section{Introduction}
\label{sect_intro}

The driving mechanism of mass outflows in Mira variables 
has not yet been fully understood.  
Recent progress in optical and infrared interferometric techniques 
has been contributing 
to studies of the region between the top of the photosphere and 
the innermost region of the circumstellar dust shell, exactly 
where mass outflows are expected to be initiated.  
Near-infrared interferometric observations of several oxygen-rich 
and S-type (C/O $\approx 1$) Mira variables 
(Mennesson et al. \cite{mennesson02}; 
Perrin et al. \cite{perrin04}; Ireland et al. \cite{ireland04}; 
Woodruff et al. \cite{woodruff04}; Fedele et al. \cite{fedele05})  
have turned out to be consistent with the presence of 
dense molecular layers extending to $\sim$2~\mbox{$R_{\star}$}, which 
were introduced to explain spectroscopic observations 
(e.g., Hinkle et al. \cite{hinkle79}; Tsuji et al. \cite{tsuji97}; 
Yamamura et al. \cite{yamamura99}).  
Furthermore, based on semi-empirical models, 
Ohnaka (\cite{ohnaka04a}) quantitatively shows that 
near-infrared ($K$ and $L^{\prime}$ bands) and mid-infrared 
(11~\mbox{$\mu$m}) interferometric and spectroscopic observations of 
three oxygen-rich and S-type Miras ($o$~Cet, R~Leo, and $\chi$~Cyg) 
can be explained by dense warm water vapor layers 
extending to $\sim$2.5~\mbox{$R_{\star}$}\ with temperatures of 1200--2000~K 
and \mbox{H$_2$O}\ column densities of the order of $10^{21}$~\mbox{cm$^{-2}$}.  
These dense water vapor layers are responsible for 
the increase of the angular size from the near-infrared to the 
mid-infrared observed toward oxygen-rich Miras.  

However, these studies have been limited to oxygen-rich or S-type 
Mira variables up to now, and the physical properties of the outer 
atmosphere of carbon stars and their temporal variations 
have not yet been well probed.  Observations of carbon-rich Miras 
as well as non-Mira carbon stars using the Short Wavelength 
Spectrometer (SWS) onboard the Infrared Space 
Observatory (ISO) revealed that while the spectra of carbon 
stars shortward of $\sim$5~\mbox{$\mu$m}\ can be fairly reproduced 
by non-gray hydrostatic or dynamical model atmospheres, 
these models entirely fail to explain the observed spectra 
longward of $\sim$5~\mbox{$\mu$m}, as demonstrated by 
J\o rgensen et al. (\cite{jorgensen00}) and Gautchy-Loidl et 
al. (\cite{gautchy-loidl04}).  These models predict very 
strong absorption due to \mbox{C$_2$H$_2$}\ and HCN at 7 and 14~\mbox{$\mu$m}, 
but the observed ISO spectra of carbon stars show only 
weak absorption due to these molecular species.  
As a possible solution for this discrepancy, 
Gautchy-Loidl et al. (\cite{gautchy-loidl04}) show 
that dynamical models with dust-driven stellar winds can 
provide a {\em qualitative} explanation of the absence of 
strong absorption features due to \mbox{C$_2$H$_2$}\ and HCN at least 
for stars with mass loss rates of $\sim \!\! 10^{-6}$~\mbox{$M_{\sun}$~yr$^{-1}$}, 
although these models cannot yet reproduce the observed strengths 
of the molecular features quantitatively.  
On the other hand, Aoki et al. (\cite{aoki98}; \cite{aoki99}) 
propose that emission from extended warm molecular layers 
containing \mbox{C$_2$H$_2$}\ and HCN may be responsible for the 
discrepancy between the ISO spectra and the photospheric 
models.  

Infrared interferometry provides us with an 
excellent opportunity to probe the nature of this controversial 
region --- the outer atmosphere 
and the innermost region of the dust shell of carbon stars. 
Van Belle et al. (\cite{vanbelle97}) observed two carbon-rich 
Miras U~Cyg and V~Cyg at 2--3 epochs in the $K$ band 
using the Infrared Optical Telescope Array (IOTA), but the 
accuracy of the derived angular sizes as well as the 
presence of molecular absorption, which affects the apparent 
stellar size in these broad-band observations, makes the deduction 
of temporal variations difficult.  
Thompson et al. (\cite{thompson02}) brought about 
a clearer picture of the phase dependence of the angular 
size of the carbon-rich Mira RZ~Peg 
based on multi-epoch narrow-band observations in the $K$-band 
using the Palomar Testbed Interferometer (PTI): 
the uniform-disk diameters measured over 
two pulsational cycles show a variation from 2.3 to 3.0~mas.   
Their measurements also show that the observed uniform-disk diameter 
monotonically increases from 2.0 to 2.4~\mbox{$\mu$m}, which can be 
interpreted as that the strong CO first 
overtone bands longward of 2.3~\mbox{$\mu$m}\ make the star appear larger 
than at shorter wavelengths where only relatively weak (but crowded) 
CN lines are present.  

However, mid-infrared interferometric observations of 
carbon stars (particularly, not very dusty ones) are still 
very scarce, let alone multi-epoch 
measurements to follow temporal variations of the physical 
properties of the photosphere and the circumstellar environment.  
The MID-infrared Interferometric instrument (MIDI) at the ESO's 
Very Large Telescope Interferometer (VLTI) 
is well suited for a study of the circumstellar environment close 
to the star.  It enables us to obtain visibilities 
with a spectral resolution of 30 or 230 from 8 to 13~\mbox{$\mu$m}, 
where the absorption features due to \mbox{C$_2$H$_2$}\ and HCN as well as 
the SiC and (featureless) amorphous carbon dust emission are 
observed for carbon stars.  

\begin{figure*}[!hbt]
\sidecaption
\includegraphics[width=12cm]{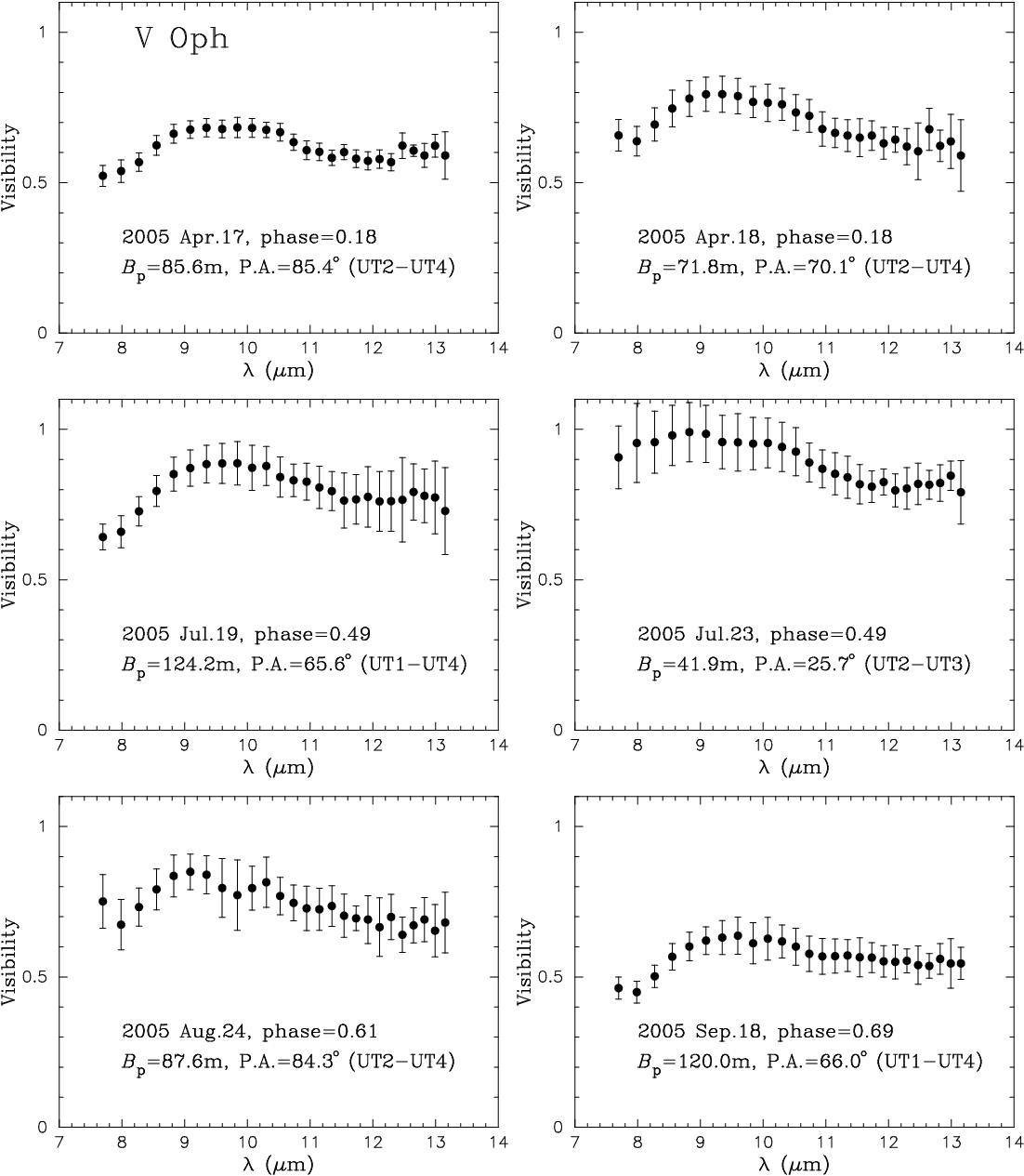}
\caption{$N$-band visibilities of V~Oph observed with MIDI 
on six nights between April and September 2005.  
}
\label{vis_obs}
\end{figure*}

In this paper, we present the first multi-epoch $N$-band 
spectro-interferometric observations of the carbon-rich Mira star 
V~Oph using MIDI.  Our object V~Oph has a period 
of 297~days (Samus et al. \cite{samus04}), and 
the $V$ magnitude varies from 7.3 to 11.6 ($F_{12\mu{\rm m}} \approx
20$~Jy).  
Its spectral energy distribution 
(SED) suggests that the circumstellar dust shell is optically thin 
with quite a low mass loss rate of $\sim \!\! 10^{-8}$~\mbox{$M_{\sun}$~yr$^{-1}$}\ 
(Groenewegen et al. \cite{groenewegen98}).  
The absence of a thick dust shell makes V~Oph suitable for 
probing the presence or absence of extended molecular layers 
in carbon-rich Miras.

\section{MIDI observations}
\label{sect_obs}

V~Oph was observed with MIDI on six nights between April 
and September 2005 in service mode (Program ID: 075-D0607, 
P.I.: K.~Ohnaka), as summarized in Table~\ref{table_obs}.  
A prism with a spectral resolution of $\lambda/\Delta \lambda \simeq 30$ 
at 10~\mbox{$\mu$m}\ was used to obtain spectrally dispersed fringes 
between 8 and 13~\mbox{$\mu$m}.  
A detailed description of the observing procedure is given in 
Przygodda et al. (\cite{przygodda03}), Leinert et al.\ (\cite{leinert04}), 
and Chesneau et al. (\cite{chesneau05a}, \cite{chesneau05b}).  

We use two different MIDI data reduction packages: MIA developed 
at the Max-Planck-Institut f\"ur Astronomie and EWS developed 
at the Leiden Observatory (MIA+EWS, ver.1.4)\footnote{Available 
at http://www.mpia-hd.mpg.de/MIDISOFT/ and 
http://www.strw.leidenuniv.nl/\textasciitilde nevec/MIDI/index.html}.   
While the MIA package is based on 
the power spectrum analysis, which measures the total power 
of observed fringes (Leinert et al. \cite{leinert04}), 
the EWS software first corrects for optical path differences 
(instrumental as well as atmospheric delays) in each scan, and 
then, the fringes are coherently added (Jaffe \cite{jaffe04}).  
We derive the interferometer transfer function at each wavelength 
(i.e., at each spectral channel) between 8 and 13~\mbox{$\mu$m}\ by observing 
calibrators whose angular diameters are known (see 
Table~\ref{table_calib}). 
In order to calibrate the visibility of V~Oph at a given wavelength, 
we use the mean of the transfer function values derived from all the 
calibrators observed on the same night as V~Oph, and 
the errors of the calibrated visibilities are estimated from the 
standard deviation of the transfer function values at each wavelength.  
The results obtained with the MIA and EWS packages show 
good agreement, and we only present the result reduced with the EWS 
package below.  
The relative errors of the calibrated visibilities 
are typically $\pm 10$--15\% (1$\sigma$), and the error 
sources are described in Ohnaka et al. (\cite{ohnaka05}).

\begin{table}
\begin{center}
\caption {Summary of MIDI observations of V~Oph: 
night, time of observation (UTC=Coordinated Universal Time), 
airmass (AM), telescope configurations (Tel., UT=Unit Telescope), 
projected baseline length $B_{\rm p}$, 
position angle of the baseline vector on the sky (P.A.), 
and variability phase estimated from the visual light curve 
compiled by the American Association of Variable Star Observers 
(AAVSO).  
}
\begin{tabular}{l c c l c c r c}\hline
\# & Night  & $t_{\rm obs}$ & AM & Tel. & $B_{\rm p}$ & P.A. & phase \\ 
   & (2005) & (UTC)         &    &            & (m)         & (\degr)& \\ 
\hline
1        & Apr. 17 & 08:39:02  & 1.1  & UT2-4  &  85.6  &  85.4 &0.18\\
2        & Apr. 18 & 04:51:38  & 1.3  & UT2-4  &  71.8  &  70.1 &0.18\\
3        & Jul. 19 & 03:03:08  & 1.1  & UT1-4  &  124.2 &  65.6 &0.49\\
4        & Jul. 23 & 22:58:21  & 1.2  & UT2-3  &  41.9  &  25.7 &0.49\\
5        & Aug. 24 & 23:49:30  & 1.1  & UT2-4  &  87.6  &  84.3 &0.61\\
6        & Sep. 18 & 23:27:32  & 1.2  & UT1-4  &  120.0 &  66.0 &0.69\\
\hline
\label{table_obs}
\vspace*{-7mm}

\end{tabular}
\end{center}
\end{table}

\begin{figure}
\includegraphics[width=8.2cm]{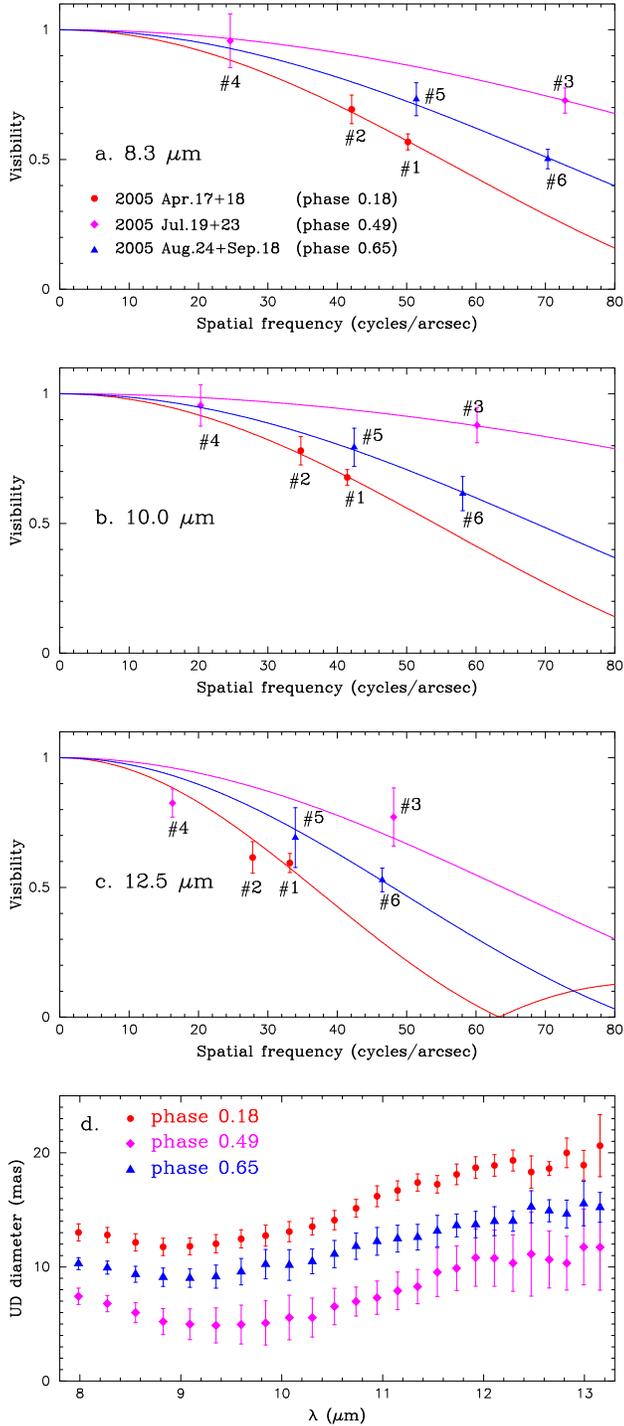}
\caption{$N$-band visibilities of V~Oph plotted as a function of 
spatial frequency at three representative wavelengths.  
{\bf a:} 8.3~\mbox{$\mu$m}, {\bf b:} 10.0~\mbox{$\mu$m}, and {\bf c:} 
12.5~\mbox{$\mu$m}. 
In these panels, the filled circles, diamonds, and triangles represent 
the data obtained at epoch 1 (phase 0.18), epoch 2 (phase 0.49), and 
epoch 3 (phase 0.65), respectively, as defined in Sect.~\ref{sect_obs}.  
The fit with uniform disks is also plotted for three epochs.  
The data set number of each visibility point as given in 
Table~\ref{table_obs} is indicated.  
{\bf d:} Uniform-disk (UD) diameters derived by fitting the visibility 
points obtained at each epoch are plotted as a function of 
wavelength.  
The filled circles, diamonds, and triangles represent 
the data obtained at epoch 1, 2, and 3, respectively.  
}
\label{visfreq_obs}
\end{figure}

Figure~\ref{vis_obs} shows the calibrated visibilities of V~Oph 
obtained on six nights (the complete set of the observed visibilities 
are electronically available in Table~4).  
The observed $N$-band visibilities except for the data \#4 taken 
with the shortest baseline on 2005 July 23 are characterized 
by a monotonic increase from 8 to $\sim$9.5~\mbox{$\mu$m}\ and a 
gradual decrease longward of $\sim$10~\mbox{$\mu$m}.  
Figures~\ref{visfreq_obs}a--c show the observed visibilities 
plotted as a function of spatial frequency at three representative 
wavelengths between 8 and 13~\mbox{$\mu$m}. 
In the modeling described below, the data sets are binned into 
three epochs: epoch 1 (2005 Apr. 17 + 18, phase 0.18), 
epoch 2 (2005 Jul. 19 + 23, phase 0.49), and epoch 3 (2005 Aug. 24 + 
Sep. 18, phase 0.61--0.69 with an average of 0.65).  
The data sets \#1 and \#5, which were obtained four months 
apart but at almost the same baseline lengths and position angles, 
reveal a temporal variation of visibility, particularly at 
8.3~\mbox{$\mu$m}. 
However, the errors of the measured visibilities become larger 
at longer wavelengths, which makes the detection of temporal 
variations less conclusive.  For example, the large errors 
of the visibilities measured at 12.5~\mbox{$\mu$m}\ hamper us from 
definitively detecting a temporal variation.  
On the other hand, 
the data sets \#3 and \#6, which were obtained two months apart 
at approximately the same baseline lengths and position angles, 
clearly indicate a temporal variation not only at 8.3~\mbox{$\mu$m}\ 
but also at 10 and 12.5~\mbox{$\mu$m}.  

\begin{table}
\begin{center}
\caption {
List of calibrators used for the calibration of the V~Oph data, 
together with 12~\mbox{$\mu$m}\ fluxes ($F_{12}$), 
uniform-disk diameters ($d_{\rm{UD}}$), 
the date as well as the time stamp ($t_{\rm obs}$) 
of MIDI observations, and airmass (AM).  
The uniform-disk diameters were taken from the CalVin 
list available at ESO
(http://www.eso.org/observing/etc/). 
The data sets used for spectrophotometric calibration of the 
spectra of V~Oph are marked with $\dagger$.  
}
\begin{tabular}{l r c c l r}\hline
Calibrator & $F_{12}$ & $d_{\rm{UD}}$ & Night  & $t_{\rm obs}$ & AM \\ 
           & (Jy)     &  (mas)        & (2005) & (UTC)         &     \\ \hline \hline
HD120404     &  13.3        & $3.03\pm 0.24$   & Apr.17  & 02:48:03 & 1.5 \\ \hline
HD142804     &  11.7         & $2.80\pm 0.08$  & Apr.17 & 04:21:08 & 1.2\\
           &               &                 & Apr.18 & 03:26:25 & 1.5\\ \hline
HD169767     &  8.8          & $2.16\pm 0.11$  & Apr.17 & 06:08:59 & 1.4\\
           &               &                 & Apr.17 & 07:18:00 & 1.2 \\
           &               &                 & Apr.17 & 07:53:19 & 1.2\\ \hline
HD150052   &  8.9  & $2.40\pm 0.07$  & Apr.17 & 08:17:33$^{\dagger}$ & 1.1\\
           &       &                 & Apr.18 & 04:29:57 & 1.4\\
           &       &                 & Jul.19 & 02:43:38$^{\dagger}$ & 1.1\\
           &       &                 & Jul.23 & 23:20:17             &1.2\\
           &       &                 & Aug.24 & 23:22:48$^{\dagger}$ &1.0\\
           &       &                 & Sep.18 & 23:51:54$^{\dagger}$ &1.3\\ \hline
HD188512     & 9.0           & $2.07\pm 0.10$  & Apr.17 & 09:04:41 & 1.3\\
           &               &                 & Jul.23 & 02:58:55 &1.3\\
           &               &                 & Jul.23 & 04:03:14 &1.2\\
           &               &                 & Jul.23 & 06:37:58 &1.4\\ \hline
HD169916   & 31.2   & $3.90\pm 0.21$  & Apr.17 & 10:08:16$^{\dagger}$ &1.0\\
           &        &                 & Apr.18 & 07:24:08 &1.1\\
           &               &                 & Jul.19 & 05:46:15 &1.2\\
           &               &                 & Jul.23 & 00:04:56$^{\dagger}$&1.3 \\
           &               &                 & Jul.23 & 05:49:35 &1.3\\ \hline
HD37160      & 9.4           & $2.08\pm 0.20$  & Apr.18 & 23:26:37&1.7\\ \hline
HD122451     & 11.4          & $0.96\pm 0.07$  & Apr.18 & 02:26:48 &1.4\\ \hline
HD152885     & 12.6          & $2.88\pm 0.09$  & Apr.18 & 05:12:20 &1.2\\
           &               &                 & Jul.23 & 00:48:11 &1.0\\ \hline
HD157236     & 10.5   & $2.58\pm 0.14$  & Apr.18 & 06:21:08$^{\dagger}$ &1.1\\
           &          &                 & Apr.18 & 08:16:35 &1.0\\
           &          &                 & Aug.24 & 01:40:04 &1.1\\ \hline
HD160668     & 8.4    & $2.30\pm 0.10$  & Apr.18 & 06:43:00 &1.1\\ \hline
HD168723     & 17.3          & $2.88\pm 0.13$  & Apr.18 & 09:17:26 &1.1\\ \hline
HD152820     & 10.9          & $2.63\pm 0.16$  & Jul.19 & 04:14:50 &1.2\\
           &         &                 & Jul.23 & 02:15:18 & 1.0\\
           &         &                 & Jul.23 & 05:01:41 & 1.5\\
           &         &                 & Sep.18 & 00:40:11$^{\dagger}$& 1.3\\ \hline
HD178345     & 12.5  & $2.50\pm 0.07$  & Aug.24 & 04:24:35 &1.3\\ \hline
HD165135     & 23.4  & $3.38\pm 0.16$  & Sep.18 & 01:30:20 &1.2 \\ \hline

\label{table_calib}
\end{tabular}
\end{center}
\end{table}

At each epoch, we fit the visibilities 
at a given wavelength with a uniform disk 
as shown in Figs.~\ref{visfreq_obs}a--c.  
While the differences in position angle 
for the data sets of the first and third epochs are not very large 
(they differ by 15.3\degr\ and 18.3\degr, respectively), 
the data sets of the 
second epoch (\#3 and \#4) were obtained with quite different position 
angles of 65.6\degr\ and 25.7\degr.  
We tentatively fit the visibility points of these data sets 
with a single uniform disk, regardless the position angles, 
because the difference in baseline length between two data sets 
makes it impossible to draw a conclusion about the possible 
presence or absence of asymmetries.  
The resulting uniform-disk diameters derived for three epochs 
are plotted as a function of wavelength in Fig.~\ref{visfreq_obs}d, 
which illustrates a temporal variation of the angular 
size of V~Oph: the object {\em appears} smaller at 
minimum light (phase 0.49) than at post-maximum (phase 0.18) or 
post-minimum (phase 0.65).  
It should be stressed, however, that 
we use uniform-disk fits to obtain some kind of representative 
angular size of the object and that it can be very misleading 
to deduce quantitatively temporal variations from these uniform-disk 
diameters, particularly if the real intensity distribution of the 
object is very different from a uniform disk.

The photospheric angular size of V~Oph 
(i.e., the stellar angular diameter measured in the continuum) 
has not been directly 
measured, but we can estimate it from the bolometric flux 
and effective temperature, if the radiation is assumed to 
be isotropic.  
The bolometric flux of V~Oph near maximum light is derived to be 
$4.4 \times 10^{-10}$~W~m$^{-2}$ from photometric data available 
in the literature, while the effective temperature is 
estimated to be 2600--3000~K (see Sect.~\ref{sect_dust_model} 
for discussion on photometric data and effective temperature).  
These bolometric flux and effective temperatures 
translate into a photospheric angular size of 4--6~mas.  
It means that the $N$-band angular sizes obtained at three epochs 
are significantly larger than the photospheric size in most cases.  
Only the 9--10~\mbox{$\mu$m}\ uniform-disk diameters obtained at 
phase 0.49 are rather close to the photospheric size.  
Therefore, for the carbon-rich Mira V~Oph, we detect the 
same trend as found in oxygen-rich Mira stars: the mid-infrared  
angular sizes are much larger than the photospheric angular 
size.  

\begin{figure}
\resizebox{\hsize}{!}{\rotatebox{0}{\includegraphics{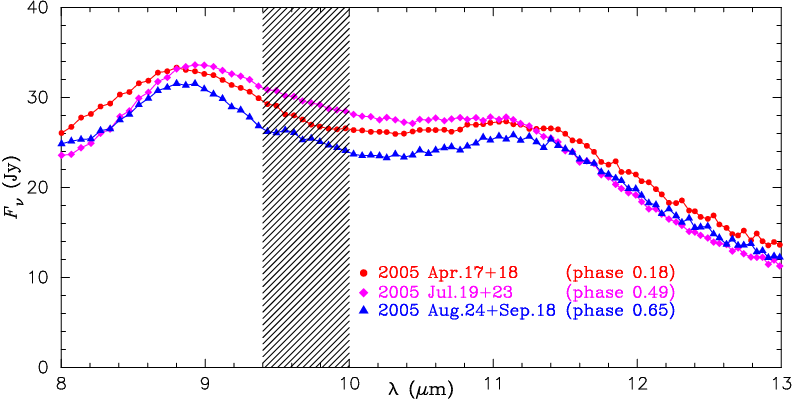}}}
\caption{$N$-band spectra of V~Oph obtained at three epochs. 
The filled circles, diamonds, and triangles represent 
the spectra obtained at epochs 1, 2, and 3, respectively.  
The shaded region represents the wavelength range severely affected 
by the telluric ozone absorption centered at 9.6~\mbox{$\mu$m}.  
}
\label{spec_obs}
\end{figure}

We also extract the absolutely calibrated $N$-band spectra of 
V~Oph from the MIDI data 
by dividing the observed spectra of V~Oph by that of a
spectrophotometric standard star and then multiplying the 
absolutely calibrated spectrum of this latter standard star.  
Interferometric calibrators observed close in time to V~Oph 
at similar airmasses are used for extracting the spectra of V~Oph as 
marked with daggers in Table~\ref{table_calib}.  
The absolutely calibrated spectra of these spectrophotometric 
standard stars are taken from Cohen et al. (\cite{cohen99}) 
if available.  Otherwise, the absolutely calibrated spectrum 
of a standard star is calculated as follows.  
First, we select a star in Cohen et al. (\cite{cohen99}) 
with the same spectral type and luminosity class as our standard 
star.  Then the flux level of the spectrum of this ``template'' 
star is scaled using its angular diameter given in 
Cohen et al. (\cite{cohen99}) and that of our standard star 
given in Table~\ref{table_calib}, and the resulting spectrum 
is used for spectrophotometric calibration.  
Finally, we average all the resulting spectra obtained at each 
epoch.  
The absolutely calibrated spectra of V~Oph observed at 
three epochs are shown in Fig.~\ref{spec_obs}. 
The errors of the MIDI spectra are difficult to estimate 
properly from the present data sets, because the number of 
standard stars observed close in time to V~Oph and at similar 
airmasses is quite limited (3, 2, and 3 standard stars at 
epoch 1, 2, and 3, respectively).  However, we estimate 
the errors of the absolutely calibrated spectra to be 10--20\%, 
based on other studies which derived $N$-band spectra 
from MIDI data (e.g., Kervella et al. \cite{kervella06}; 
Poncelet et al. \cite{poncelet06}; Quanz et al. \cite{quanz06}).

Figure~\ref{spec_obs} reveals that the spectra are characterized by 
absorption between 8 and 9~\mbox{$\mu$m}\ and a broad emission 
feature centered at $\sim$11.3~\mbox{$\mu$m}.  
The feature seen between 8 and 9~\mbox{$\mu$m}\ is actually 
the tail of the broad absorption feature centered at $\sim$7.5~\mbox{$\mu$m}\ 
due to the \mbox{C$_2$H$_2$}\ \mbox{$\nu_{4}$+$\nu_{5}$}\ band and 
the HCN 2$\nu_{2}$ band.  
However, as shown in Sect.~\ref{subsect_moldata}, 
the HCN 2$\nu_{2}$ band extends only up to $\sim$8.3~\mbox{$\mu$m}, 
which leaves \mbox{C$_2$H$_2$}\ as the dominant opacity source between 8 and 
9~\mbox{$\mu$m}.  The broad emission feature 
at $\sim$11.3~\mbox{$\mu$m}\ is attributed to SiC dust.  
In addition to these conspicuous features within the $N$ band, 
a broad absorption feature centered at $\sim$14~\mbox{$\mu$m}\ due to the 
\mbox{C$_2$H$_2$}\ \mbox{$\nu_{5}$}\ and HCN $\nu_{2}$ bands is identified in ISO 
spectra of other carbon-rich Mira variables (e.g., 
Yamamura et al. \cite{yamamura97}; Hron et al. \cite{hron98}; 
Aoki et al. \cite{aoki99}).  
This absorption feature extends down to $\sim$11~\mbox{$\mu$m}, where 
it becomes blended with the SiC emission feature.  
Comparison between the spectra at phases 0.18 and 0.49 reveals 
that the 8~\mbox{$\mu$m}\ absorption feature becomes somewhat stronger at 
minimum light than at post-maximum. 
This trend is similar to that found in the Mira-like carbon star 
R~Scl by Hron et al. (\cite{hron98}).  
We will discuss this variation of the 8~\mbox{$\mu$m}\ feature 
in Sect.~\ref{subsect_wme_model}.

\section{Modeling with a dust shell}
\label{sect_dust_model}

As briefly discussed in Sect.~\ref{sect_obs}, the $N$-band 
uniform-disk diameters measured with MIDI are remarkably 
larger than the estimated photospheric angular size, 
which suggests the contribution of circumstellar material 
in the mid-infrared angular size.  This contribution 
can be of the molecular and/or dust origin.  We first attempt 
to interpret the observed $N$-band visibilities and spectra 
with dust shell models.  

\begin{figure*}[hbt]
\sidecaption
\includegraphics[width=12cm]{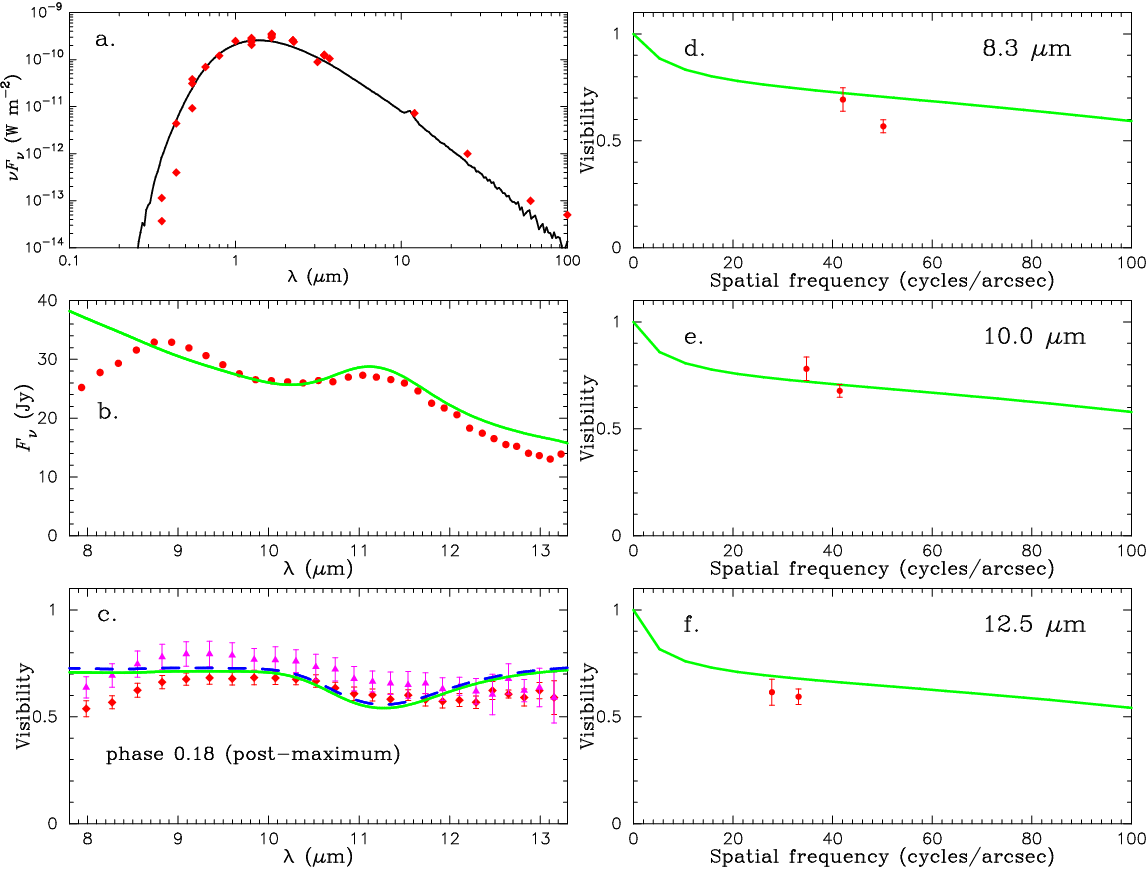}
\caption{Dust shell model containing amorphous 
carbon and SiC which well reproduces the observed SED for 
phases near maximum light.  The dust shell parameters are 
\mbox{$r_{\rm in}$}\ = 2.5~\mbox{$R_{\star}$}, 
\mbox{$\tau_{V}^{\rm amc}$}\ = 0.6, and \mbox{$\tau_{V}^{\rm SiC}$}\ = 0.26.  
{\bf a:} Comparison of the observed and model SEDs.  
The observed SEDs are represented with the filled diamonds, 
while the model SED is represented with the solid line. 
{\bf b:} Comparison of the observed and model $N$-band spectra. 
The filled circles represent the MIDI spectrum obtained at 
phase 0.18, while the solid line represents the model prediction. 
{\bf c:} The filled diamonds and triangles represent the $N$-band 
visibilities measured with projected baseline lengths of 85.6 and 
71.8~m, respectively, while the solid and dashed lines represent 
the corresponding model visibilities.  
{\bf d}--{\bf f:} The solid lines represent the model visibilities 
at 8.3, 10.0, and 12.5~\mbox{$\mu$m}\ plotted as a function of spatial 
frequency.  The observed data are plotted with the filled circles. 
}
\label{dust_model}
\end{figure*}

We calculate spherically symmetric dust shell models, 
using the multi-dimensional Monte Carlo code \mbox{\sf mcsim\_mpi}\ 
(Ohnaka et al. \cite{ohnaka06a}), which computes dust 
temperatures in radiative equilibrium and generates model SEDs.  
Then the model image at each wavelength within 
the $N$ band is produced with the ray-trace method as described 
in Ohnaka et al. (\cite{ohnaka06a}), and the corresponding 
monochromatic visibility 
is calculated by taking the Hankel transform of the model image.  
The monochromatic visibilities are then spectrally convolved 
to a resolution of 30 to match the MIDI observations.  
The effective temperature of V~Oph derived by various authors 
ranges from 2650 to 3010~K (Ohnaka \& Tsuji \cite{ohnaka96}; 
Bergeat et al. \cite{bergeat01}; Groenewegen \cite{groenewegen06}), 
but the phase dependence of effective 
temperature of V~Oph is not well established.  Therefore, 
in the dust shell modeling below, we adopt a fixed effective 
temperature of 2800~K, which is roughly the mean of the above range, 
and the central star is approximated with the blackbody of 
2800~K.  
We assume that the dust shell consists of amorphous carbon and 
SiC, following the results of Groenewegen et al. 
(\cite{groenewegen98}) and Groenewegen (\cite{groenewegen06}),  
who concluded that the dust chemical composition around V~Oph is 
dominated by amorphous carbon (85\%) with a minor contribution of 
SiC (15\%).  
We compute a grid of models with the following parameter 
ranges: \mbox{$r_{\rm in}$}\ (\mbox{$R_{\star}$}) = 2.5 ... 5.0 
($\Delta \mbox{$r_{\rm in}$}\ = 0.5$), 
\mbox{$\tau_{V}^{\rm amc}$}\ = 0.2 ... 0.8 
($\Delta \mbox{$\tau_{V}^{\rm amc}$} = 0.2$), where 
\mbox{$r_{\rm in}$}\ and \mbox{$\tau_{V}^{\rm amc}$}\ denote 
the inner boundary radius 
(assumed to be the same for amorphous carbon and SiC) and 
the optical depth of amorphous carbon dust at 0.55~\mbox{$\mu$m}, 
respectively.  
The relative contribution of SiC, 
\mbox{$\tau_{V}^{\rm SiC}$}/(\mbox{$\tau_{V}^{\rm amc}$} + \mbox{$\tau_{V}^{\rm SiC}$}) is 
varied from 0.1 to 0.4 with an increment of 0.1, where \mbox{$\tau_{V}^{\rm SiC}$}\ 
denotes the optical depth of SiC dust at 0.55~\mbox{$\mu$m}.  
The opacities of 
amorphous carbon and SiC are calculated from complex refractive 
indices presented by Rouleau \& Martin (\cite{rouleau91}, AC1 sample) 
and by P\'{e}gouri\'{e} (\cite{pegourie88}), respectively, assuming 
a single grain size of 0.1~\mbox{$\mu$m}\ for both species.  
The grain density distribution is assumed to be proportional to 
$r^{-2}$, and the outer radius of the shell 
is fixed to $10^3 \times \mbox{$r_{\rm in}$}$.

Figure~\ref{dust_model} shows a comparison between the observational 
data and a model which can well reproduce the SED observed near 
maximum light.  
We derived the observed SED from optical and infrared photometric data 
available in the literature (Landolt \cite{landolt68}, 
\cite{landolt73}; Walker \cite{walker79}; 
Noguchi et al. \cite{noguchi81}; Whitelock et al. \cite{whitelock00}; 
IRAS Point Source Catalog) and dereddened with $A_V$ = 0.6 
(Groenewegen et al. \cite{groenewegen98}) 
using the method of Savage \& Mathis (\cite{savage79}) with 
$A_{V} = 3.1 E(B - V)$.  
Unfortunately, most of the photometric data, particularly in the 
optical, were obtained at 
phases near maximum light, and therefore, we can compare 
model SEDs with the observed one only for phase 0.18.  
As can be seen in Figs.~\ref{dust_model}a and \ref{dust_model}b, 
the SED at maximum light 
as well as the MIDI spectrum obtained at phase 0.18 
can be explained by  the model with \mbox{$r_{\rm in}$}\ = 2.5~\mbox{$R_{\star}$}, 
\mbox{$\tau_{V}^{\rm amc}$}\ = 0.6 (0.0038 at 11.3~\mbox{$\mu$m}), 
and \mbox{$\tau_{V}^{\rm SiC}$}\ = 0.26 (0.0078 at 11.3~\mbox{$\mu$m}).  
The temperatures of amorphous carbon and SiC dust at the inner 
boundary are $\sim$1580~K, which roughly corresponds to the 
condensation temperature of amorphous carbon predicted by 
dynamical models (e.g., Nowotny et al. \cite{nowotny05a}) and 
is in agreement 
with the value that Groenewegen et al. (\cite{groenewegen98}) 
derived from a SED-fitting (note, however, that condensation temperatures 
generally depend on detailed grain nucleation and growth processes).  
The model cannot reproduce the 8~\mbox{$\mu$m}\ feature, because the 
star is approximated with a blackbody without a detailed model 
atmosphere.  
This is also the reason the model overestimates the flux in the 
optical where strong molecular absorption is present.  
Comparison of the visibilities shown in Fig.~\ref{dust_model}c 
reveals that the dust shell model can roughly reproduce the observed 
visibility levels.  
The dip at $\sim$11.3~\mbox{$\mu$m}\ seen in the model visibilities 
corresponds to the position of the SiC emission feature.  
The flux contribution of the dust shell becomes larger 
due to the SiC emission feature at these wavelengths, which 
leads to lower visibilities.  
It should be noted that the scaling factor between a model SED 
and the observed one is given by $(\mbox{$R_{\star}$}/d)^2$, where $d$ is the 
distance of V~Oph.  The model visibilities are scaled 
with the angular radius of the central star ($\mbox{$R_{\star}$}/d$) 
derived by fitting the model SEDs to the observed data.  

However, there are noticeable discrepancies between the 
model visibilities and the MIDI data.  
Firstly, the models cannot reproduce the observed wavelength 
dependence of visibility between 8 and 9~\mbox{$\mu$m}\ as well as 
longward of 12~\mbox{$\mu$m}, where the \mbox{C$_2$H$_2$}\ (and HCN) 
absorption features are present, although the error bars of the 
observed visibilities are rather large in the latter wavelength 
range.  The fact that the observed visibilities are smaller than 
the model at 8--9~\mbox{$\mu$m}\ and $\ga$12~\mbox{$\mu$m}\ suggests that 
the observed angular sizes are larger than that predicted by 
the dust shell model.  
Secondly, the models cannot reproduce 
the shape of the observed visibilities as a function of spatial 
frequency, which can be seen in Fig.~\ref{dust_model}d: 
the model visibility at 8.3~\mbox{$\mu$m}\ 
is somewhat too flat compared to those observed.  
Since the dust shell is over-resolved with the baselines used in 
our MIDI observations (the dust shell visibility component is 
seen as steep visibility drops at spatial frequencies from 
0 to $\sim$5~cycles/arcsec in Figs.~\ref{dust_model}d--f), 
the visibility shapes measured by our 
observations correspond to the central star component.  
Therefore, this disagreement of visibility shape means that 
the observed $N$-band angular size of the central star is larger 
than that used in the model.  However, the angular size of the 
central star is derived by fitting the model SED to the observed one, 
and therefore, a mere significant change of the angular size of 
the central star 
would lead to a poorer match to the observed SED including the 
$N$-band spectrum.  
Changes in effective temperature, dust chemistry, and 
density gradient cannot 
explain the observed visibility shapes, either.  
Therefore, it can be postulated that there is some additional 
component close to the star, which makes the central source 
{\em appear} larger than the star itself.

\section{Modeling with C$_{\mbox{\small 2}}$H$_{\mbox{\small 2}}$ layers 
 and a dust shell}
\label{sect_moldust}

As in oxygen-rich Mira variables which have dense \mbox{H$_2$O}\ layers 
extending to 2--3~\mbox{$R_{\star}$}, it is possible that similar layers 
consisting of polyatomic molecules exist around carbon-rich Miras, 
which makes the central star appear larger than its photospheric size. 
In carbon-rich atmospheres, \mbox{C$_2$H$_2$}\ and HCN are the most abundant 
polyatomic species with spectral features in the $N$ band, 
which makes them the most likely candidates responsible for the 
larger apparent size of the central star component.  
In this section, we attempt to explain the observed data with 
models consisting of such molecular layers and an optically thin 
dust shell.

\subsection{Molecular data}
\label{subsect_moldata}

First, we describe molecular line opacity data 
necessary for computing models with \mbox{C$_2$H$_2$}\ (+ HCN) layers.  
In the wavelength region between 8 and 9~\mbox{$\mu$m}, 
the \mbox{C$_2$H$_2$}\ \mbox{$\nu_{4}$+$\nu_{5}$}\ band is the 
dominant opacity source.  
It is also necessary to take the \mbox{C$_2$H$_2$}\ \mbox{$\nu_{5}$}\ 
band centered at 
14~\mbox{$\mu$m}\ into account for modeling the $N$-band visibilities and 
spectra, because this band extends to $\sim$11~\mbox{$\mu$m}.  
The HCN 2$\nu_{2}$ band is centered at $\sim$7~\mbox{$\mu$m}, and its 
tail extends to $\sim$8.3~\mbox{$\mu$m}.  
The HCN $\nu_{2}$ band, which is observed together with the 
\mbox{C$_2$H$_2$}\ \mbox{$\nu_{5}$}\ band in other carbon-rich Miras, 
is centered at 
14~\mbox{$\mu$m}\ and extends to $\sim$11~\mbox{$\mu$m}.  
Therefore, the contribution of HCN appears mainly at wavelengths 
between 11 and 13~\mbox{$\mu$m}, blended with 
the \mbox{C$_2$H$_2$}\ \mbox{$\nu_{5}$}\ band and the SiC dust feature.  

Despite its importance in various astrophysical environments, the 
line list of \mbox{C$_2$H$_2$}\ is far from complete, particularly for high 
temperatures relevant to stellar photospheres and outer atmospheres 
(temperatures of 1000--1600~K are derived in our modeling below). 
There are a number of laboratory measurements of the molecular 
constants of the \mbox{$\nu_{4}$+$\nu_{5}$}\ and \mbox{$\nu_{5}$}\ 
bands, but these experiments were performed at room temperatures.  
For example, Kabbadj et al. (\cite{kabbadj91}) measured the 
molecular constants up to $V_4 + V_5 \le 4$, which means that 
only 10 and 6 transitions among different ($V_4$, $V_5$) are 
included for the \mbox{$\nu_{5}$}\ and \mbox{$\nu_{4}$+$\nu_{5}$}\ 
bands, respectively.  
Therefore, many hot bands expected in stellar and circumstellar 
environments are not included, and these lists are not 
appropriate for our modeling.  
On the other hand, Tsuji (\cite{tsuji84}) presents a totally different 
approach to deal with \mbox{C$_2$H$_2$}\ opacity.  He makes use of the band model 
method, which analytically calculates the absorption cross 
section at a given wavelength from basic molecular constants, 
instead of a detailed line list.  Particularly, if the molecule 
is assumed to be a rigid-rotator-harmonic-oscillator (RRHO), 
calculations of absorption cross sections can be easily 
performed\footnote{Figure~3 of Tsuji (\cite{tsuji84}), which 
shows the absorption cross section of the \mbox{C$_2$H$_2$}\ \mbox{$\nu_{5}$}\ band,  
has an error in the numerical labels of the ordinate.  The 
correct range of the ordinate is from $-22$ to $-17$ instead of $-23$ 
to $-18$.  However, this is a mere graphical error, and the results of 
the calculations in Tsuji (\cite{tsuji84}) were performed with the 
correct cross sections (Tsuji, priv.~comm.).}.  
In the present work, we use this band model opacity computed in the 
RRHO approximation to represent the \mbox{C$_2$H$_2$}\ opacity, because neither 
laboratory measurements nor ab initio calculations at high temperatures 
are yet available for \mbox{C$_2$H$_2$}\ lines, and individual lines are not 
resolved with the MIDI's spectral resolution of 30.  

\begin{figure}
\resizebox{\hsize}{!}{\rotatebox{0}{\includegraphics{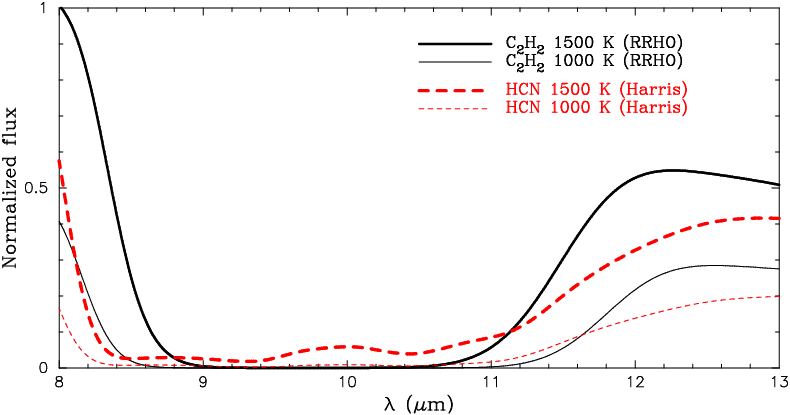}}}
\caption{Emission spectra of \mbox{C$_2$H$_2$}\ and HCN from a slab of 1000 and 
1500~K.  The column densities of \mbox{C$_2$H$_2$}\ and HCN are set to 
$10^{20}$~\mbox{cm$^{-2}$}.  The spectra are convolved with a spectral 
resolution of 30.  The opacity of \mbox{C$_2$H$_2$}\ is computed with the band 
model method as described in Sect.~\ref{subsect_moldata}, while that of 
HCN is computed from the line list of 
Harris et al. (\cite{harris02}).  
}
\label{emsflux}
\end{figure}

Compared to \mbox{C$_2$H$_2$}, studies on HCN line lists including high 
vibrational levels are more advanced.  
Recently, Harris et al. (\cite{harris02}) have calculated a 
new, extensive ab initio line list of HCN/HNC including 
energy levels up to 18\,000~\mbox{cm$^{-1}$}.  We use their line list 
to calculate the HCN opacity in the modeling below. 
Figure~\ref{emsflux} shows emission spectra of \mbox{C$_2$H$_2$}\ and 
HCN from a slab of 1000 and 1500~K with the column densities 
of both molecules set to $10^{20}$~\mbox{cm$^{-2}$}.  The HCN opacity 
was calculated assuming local thermodynamical equilibrium 
(LTE) adopting a Gaussian profile with a FWHM of 5~\mbox{km s$^{-1}$}, 
and the opacity of \mbox{C$_2$H$_2$}\ was calculated with the band 
model method with the RRHO approximation, which also implicitly 
assumes LTE.  
The spectra 
are convolved with a spectral resolution of 30 as used in the 
MIDI observations.  
The figure illustrates that \mbox{C$_2$H$_2$}\ is the dominant opacity source 
at wavelengths between 8 and 9~\mbox{$\mu$m}\ as well as longward of 
11~\mbox{$\mu$m}, while the contribution of HCN appears primarily at 
wavelengths longer than 11~\mbox{$\mu$m}\ (and at the shorter edge 
of the $N$ band).  
Since this contribution of HCN is blended with the 
\mbox{C$_2$H$_2$}\ \mbox{$\nu_{5}$}\ 
band and the SiC dust feature, the effects of HCN on spectra 
and visibilities tend to be masked by the \mbox{C$_2$H$_2$}\ and SiC 
features, which hinders us from putting strong constraints on 
the HCN column density.  
Therefore, in the models presented below, we first consider 
\mbox{C$_2$H$_2$}\ alone and examine the effects of HCN later.  

\subsection{Result of modeling with \mbox{C$_2$H$_2$}\ layers and a dust shell}
\label{subsect_wme_model}

\begin{figure*}
\resizebox{\hsize}{!}{\rotatebox{0}{\includegraphics{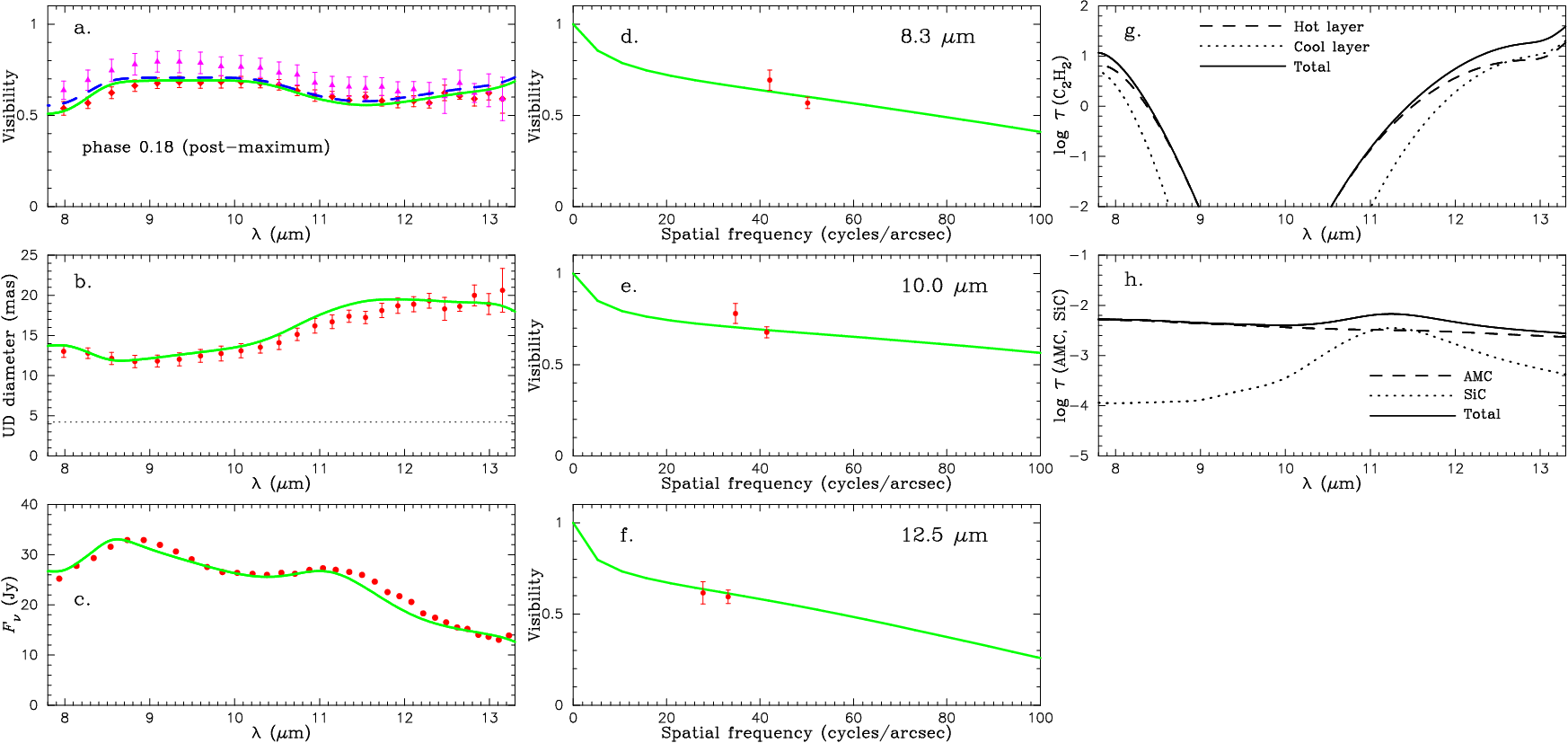}}}
\caption{Comparison between observational data at phase 
0.18 (post-maximum) and the best-fit model consisting of 
\mbox{C$_2$H$_2$}\ layers and a dust shell.   
The model parameters are given in Table~\ref{wme_param_table}.
{\bf a:} The filled diamonds and triangles represent the $N$-band 
visibilities measured at phase 0.18 
with projected baseline lengths of 85.6 and 
71.8~m, respectively, while the solid and dashed lines represent 
the corresponding model visibilities.  
{\bf b:} The filled circles represent the uniform-disk 
diameters observed at phase 0.18, while the solid line represents 
the model prediction, which is derived by fitting the model 
visibilities predicted for baselines of 85.6 and 71.8~m 
with a uniform disk at each wavelength.  
The dotted line represents the photospheric angular diameter 
derived by fitting the model $N$-band spectrum to the observed data.  
{\bf c:} Comparison of the observed and model $N$-band spectra. 
The filled circles represent the MIDI spectrum obtained at 
phase 0.18, while the solid line represents the model prediction. 
{\bf d}--{\bf f:} The solid lines represent the model visibilities 
at 8.3, 10.0, and 12.5~\mbox{$\mu$m}\ plotted as a function of spatial 
frequency.  The observed data are plotted with the filled circles. 
{\bf g:} The \mbox{C$_2$H$_2$}\ optical depths of the hot and cool layers 
(convolved to a spectral resolution of 30).  
{\bf h:} The optical depths for amorphous carbon (AMC) and 
SiC dust (convolved to a spectral resolution of 30).  
}
\label{wme_model_max}
\end{figure*}

We start from models consisting of one \mbox{C$_2$H$_2$}\ layer with a constant 
temperature and density and an optically thin dust shell of amorphous 
carbon and SiC, similar to the model used by 
Ohnaka et al. (\cite{ohnaka05}).  
In this model, the temperature, 
outer radius, and the \mbox{C$_2$H$_2$}\ column density of the layer are varied 
as free parameters.  In addition, the inner radius of this layer 
is either set to be equal to \mbox{$R_{\star}$}\ or varied as a free parameter 
to model a detached \mbox{C$_2$H$_2$}\ layer.  
The central star itself is approximated with a blackbody of 
the effective temperature \mbox{$T_{\rm eff}$}, which is fixed to 2800~K as in 
the dust shell models discussed above.  
Using the \mbox{C$_2$H$_2$}\ opacity given above, the monochromatic intensity 
distribution of the \mbox{C$_2$H$_2$}\ layer and the star is calculated as 
described in Ohnaka et al. (\cite{ohnaka04b}).  
The intensity profile of the dust shell (excluding the central star) 
at each wavelength in the $N$ band is calculated with the \mbox{\sf mcsim\_mpi}\ 
code and ray-tracing method as in the above dust shell models, 
and this intensity profile is added 
to that resulting from the central star and the \mbox{C$_2$H$_2$}\ layer.  
The inner boundary radius of the dust shell is set to 2.5~\mbox{$R_{\star}$}, 
because the temperature at this radius (1500--1600~K) 
approximately corresponds to the condensation temperature 
of amorphous carbon grains.  
The spectrally convolved $N$-band visibilities are then computed as 
outlined in Sect.~\ref{sect_dust_model}.  
However, it has turned out that these models cannot 
reproduce the shape of the 8~\mbox{$\mu$m}\ absorption feature and the 
observed $N$-band visibilities simultaneously for the following 
reason.  In these one-layer models, the \mbox{C$_2$H$_2$}\ layer cannot be too 
extended, because otherwise the \mbox{$\nu_{4}$+$\nu_{5}$}\ and \mbox{$\nu_{5}$}\ bands 
would appear in emission, in disagreement with the observations.  
However, with a modest geometrical extension of the \mbox{C$_2$H$_2$}\ layer, 
the angular sizes observed between 8 and 9~\mbox{$\mu$m}, which are 
remarkably larger than the photospheric size, cannot be reproduced.

\begin{figure*}
\resizebox{\hsize}{!}{\rotatebox{0}{\includegraphics{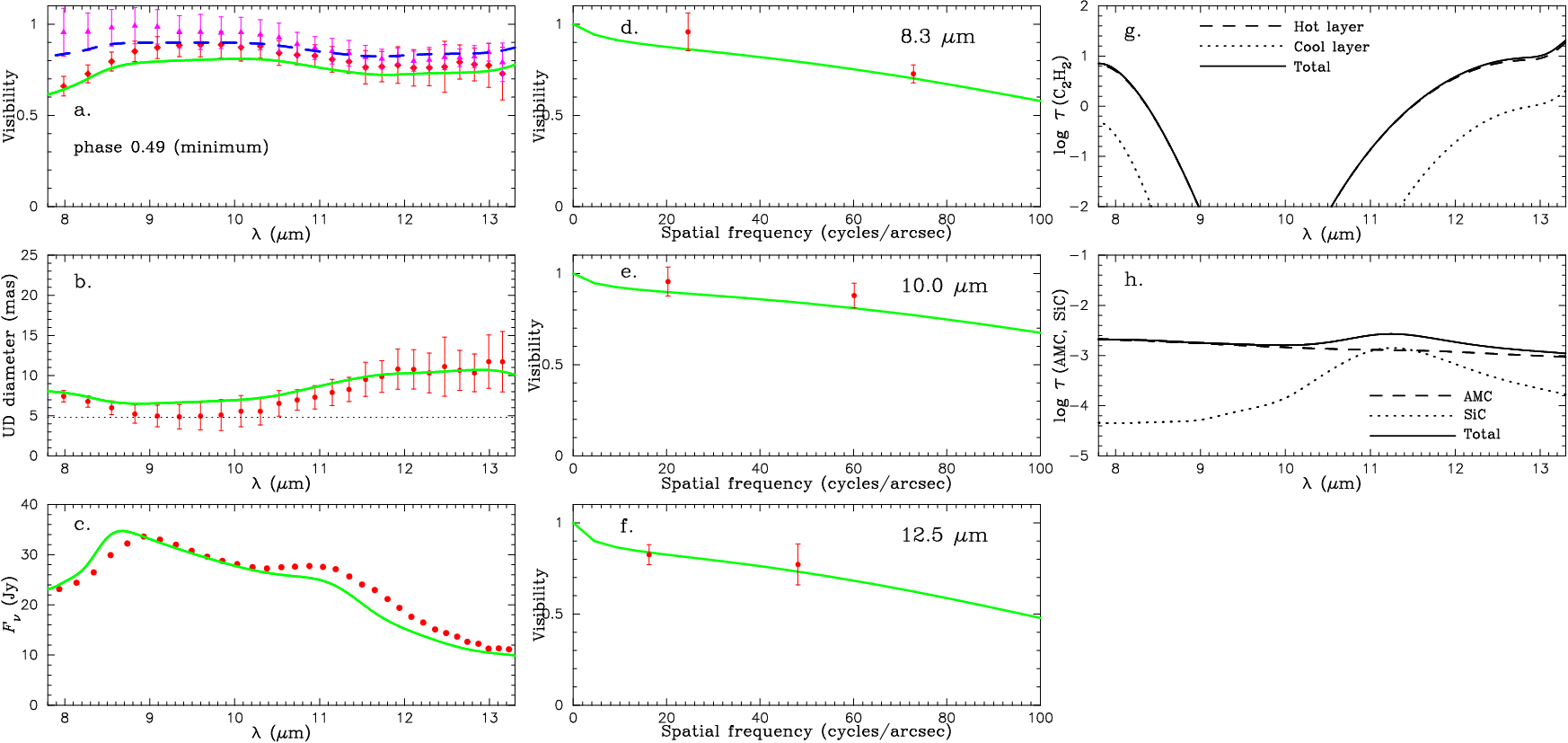}}}
\caption{
Same as Fig.~\ref{wme_model_max}, but for phase 0.49 (minimum 
light).  In the panel {\bf a}, the filled diamonds and triangles 
represent the $N$-band visibilities measured 
with projected baseline lengths of 124.2 and 
41.9~m, respectively, while the solid and dashed lines represent 
the corresponding model visibilities.
In the panel {\bf b}, the solid line represents 
the model prediction, which is derived by fitting the model 
visibilities predicted for baselines of 124.2 and 41.9~m with 
uniform disks.  
In the panel {\bf g}, the solid and dashed lines almost entirely 
overlap.  
}
\label{wme_model_mid}
\end{figure*}

Therefore, we examine models with a dust shell and two \mbox{C$_2$H$_2$}\ 
layers, which are similar to the warm water vapor envelope model 
for oxygen-rich Miras considered by Ohnaka (\cite{ohnaka04a}).  
In these models, the star is surrounded by two layers containing 
\mbox{C$_2$H$_2$}\ with 
different temperatures \mbox{$T_{\rm hot}$}\ and 
\mbox{$T_{\rm cool}$}, and the hot and cool 
layers extend to \mbox{$R_{\rm hot}$}\ and \mbox{$R_{\rm cool}$}, respectively 
(see Fig.~\ref{wme_cartoon} for illustration of the model).  
The temperature 
and the density are assumed to be constant in each of these 
layers.   The \mbox{C$_2$H$_2$}\ column densities in the hot and cool 
layers are denoted as \mbox{$N_{\rm hot}$}\ and \mbox{$N_{\rm cool}$}, respectively.  
The computation of intensity profiles from such two layers 
are described in Ohnaka (\cite{ohnaka04b}), and the total 
intensity profile is the sum of this component and 
the dust shell component with its inner boundary fixed 
to 2.5~\mbox{$R_{\star}$}.  From these intensity profiles, spectrally convolved 
visibilities are calculated as in the case of the above one-layer 
model.  
We search for a best-fit model 
within the following parameter range: 
\mbox{$T_{\rm hot}$}\ (K) = 1300 ... 1800 with 
$\Delta \mbox{$T_{\rm hot}$}$ (K) = 100, 
\mbox{$T_{\rm cool}$}\ (K) = 800 ... 1200~K 
with $\Delta \mbox{$T_{\rm cool}$}$ (K) = 100, 
\mbox{$R_{\rm hot}$}\ (\mbox{$R_{\star}$}) = 1.1 ... 1.6 
with $\Delta \mbox{$R_{\rm hot}$}$ (\mbox{$R_{\star}$}) = 0.1, 
\mbox{$R_{\rm cool}$}\ (\mbox{$R_{\star}$}) = 1.2 ... 2.0 
with $\Delta \mbox{$R_{\rm cool}$}$ (\mbox{$R_{\star}$}) = 0.1 
(with conditions $\mbox{$R_{\rm hot}$} < \mbox{$R_{\rm cool}$}$), 
$\log \mbox{$N_{\rm hot}$}$ (\mbox{cm$^{-2}$}) = 19 ... 22, 
and $\log \mbox{$N_{\rm cool}$}$ (\mbox{cm$^{-2}$}) = 
18 ... 21 with $\Delta \log \mbox{$N_{\rm hot}$} = \Delta \log \mbox{$N_{\rm cool}$}$ = 0.5. 
The optical depth of the dust shell is varied between \mbox{$\tau_{V}^{\rm amc}$}\ = 
0.1 and 0.8 with $\Delta \mbox{$\tau_{V}^{\rm amc}$}$ = 0.1, 
while the fraction of SiC, 
$f_{\rm SiC} = \mbox{$\tau_{V}^{\rm SiC}$} / (\mbox{$\tau_{V}^{\rm amc}$} + 
\mbox{$\tau_{V}^{\rm SiC}$})$, is varied from 0.1 to 0.4.

It should be noted here that not only the wavelength dependence 
of the observed visibilities but also the visibility shape as a 
function of spatial frequency constrained by the measurements at 
two different baseline lengths are crucial for this modeling.  
That is, the effect of 
the extended dust shell, which is over-resolved at the baseline 
lengths used in our MIDI observations, is to lower the total 
visibility by an amount equal to its fractional flux contribution 
at a given wavelength, as explained in Sect.~\ref{sect_dust_model}.  
Therefore, the parameters of the dust shell 
roughly defines absolute visibility levels.  
On the other hand, the presence of the \mbox{C$_2$H$_2$}\ layers makes the 
visibility (as a function of spatial frequency) steeper 
at higher spatial frequencies 
than models consisting of a bare central star and a dust shell.  
The \mbox{C$_2$H$_2$}\ layers also give rise to 
absorption or emission at wavelengths between 8 and 9~\mbox{$\mu$m}\ and 
longward of $\sim$11~\mbox{$\mu$m}, although in this latter wavelength range, 
the \mbox{C$_2$H$_2$}\ feature is blended with the SiC emission feature.  
In this manner, it is possible to constrain the parameters of 
the hot and cool \mbox{C$_2$H$_2$}\ layers as well as the dust shell.

\begin{figure*}
\resizebox{\hsize}{!}{\rotatebox{0}{\includegraphics{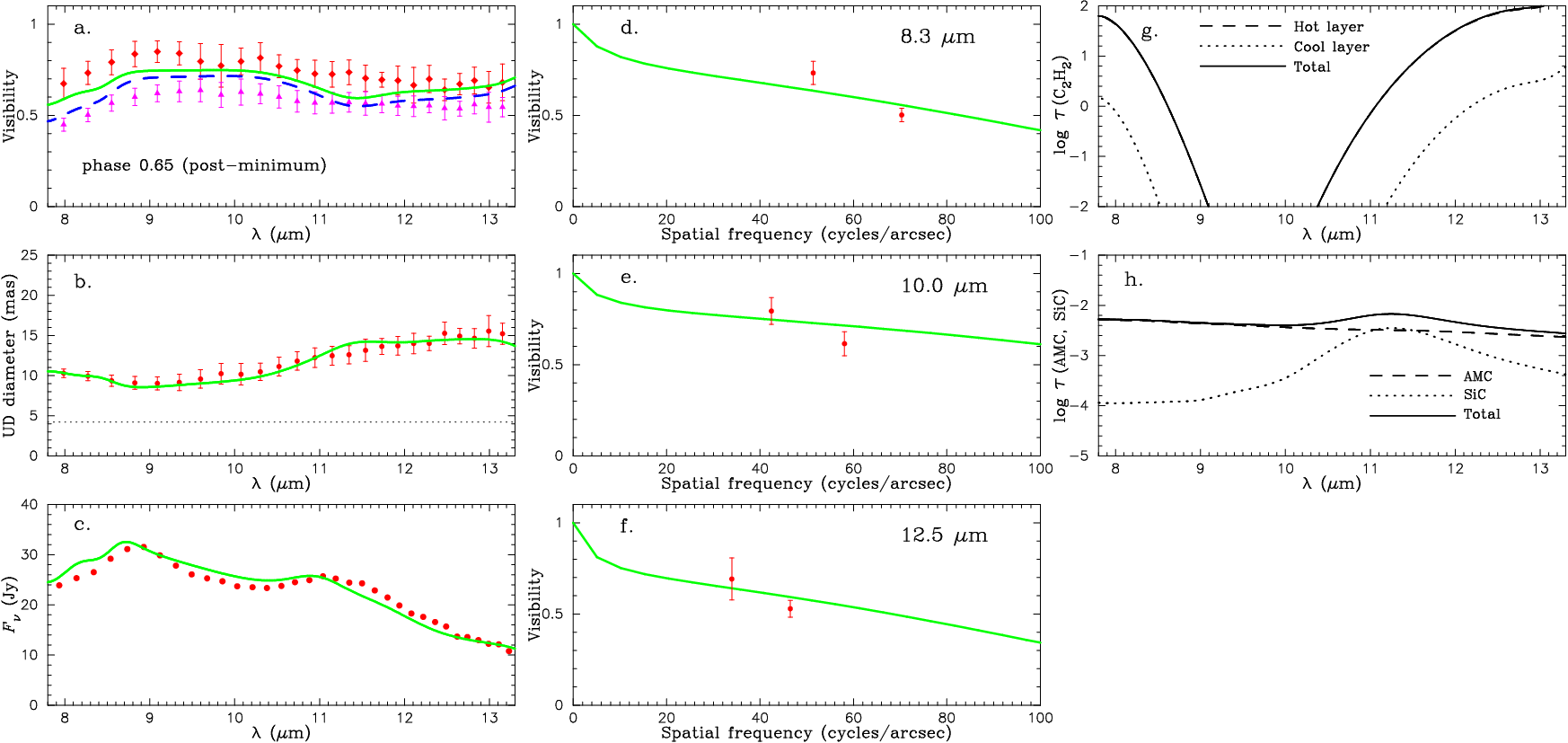}}}
\caption{
Same as Fig.~\ref{wme_model_max}, but for phase 0.65 
(post-minimum).  In the panel {\bf a}, the filled diamonds and triangles 
represent the $N$-band visibilities measured 
with projected baseline lengths of 87.6 and 
120.0~m, respectively, while the solid and dashed lines represent 
the corresponding model visibilities.
In the panel {\bf b}, the solid line represents 
the model prediction, which is derived by fitting the model 
visibilities predicted for baselines of 87.6 and 120.0~m with uniform 
disks.  In the panel {\bf g}, the solid and dashed lines almost entirely 
overlap.  
}
\label{wme_model_min}
\end{figure*}

Figures~\ref{wme_model_max}, \ref{wme_model_mid}, and 
\ref{wme_model_min} show the best-fit models consisting of 
\mbox{C$_2$H$_2$}\ layers and a dust shell for three epochs observed with MIDI, 
and the parameters of these best-fit models are given in 
Table~\ref{wme_param_table}.  The figures demonstrate 
that the models can reproduce the observed $N$-band spectra, 
particularly the absorption feature between 8 and 9~\mbox{$\mu$m}.  
At the same time, the observed $N$-band visibilities --- not only the 
absolute visibility levels but also their wavelength 
dependence --- are also fairly reproduced by these models.  
The visibility increase from 
8 to 10~\mbox{$\mu$m}, which is seen in all the data sets except for 
\#4 taken with the shortest baseline at minimum light, can be 
well explained by the models.  
The visibilities predicted at 12--13~\mbox{$\mu$m}, where the tail of the 
\mbox{C$_2$H$_2$}\ \mbox{$\nu_{5}$}\ band is located, are also in 
reasonable agreement with the MIDI observations.   
These results suggest that the optically thick emission 
($\tau_{\rm line} \approx $1--100, see 
Figs.~\ref{wme_model_max}g, \ref{wme_model_mid}g, and 
\ref{wme_model_min}g) from the \mbox{C$_2$H$_2$}\ layers, together with 
the dust emission, is responsible for the $N$-band angular size 
significantly larger than the photospheric size and that \mbox{C$_2$H$_2$}\ plays 
a role similar to \mbox{H$_2$O}\ in oxygen-rich environments.  
Also, as will be discussed below, this optically thick 
emission plays an important role in the interpretation of the 
controversial apparent depth of the 7 and 14~\mbox{$\mu$m}\ 
\mbox{C$_2$H$_2$}\ features outlined in Sect.~\ref{sect_intro}.  
We note that as in the case of the dust shell 
models discussed above, the model visibilities are scaled using the 
angular radius of the central star (stellar continuum angular radius) 
derived by fitting the model $N$-band spectra to those observed. 
The stellar photospheric angular sizes derived in this manner are also 
given in Table~\ref{wme_param_table}.  
In this method, the uncertainty of the stellar diameter is primarily 
determined by the uncertainties of the absolutely calibrated $N$-band 
spectra of $\sim$10--20\%.  Based on this, we estimate the error of 
the stellar diameter to be $\pm0.3$~mas.  
It is noteworthy that while the observed uniform-disk diameters are 
the smallest at minimum light (see Fig.~\ref{visfreq_obs}d), 
the stellar continuum diameter does not show such a phase dependence 
(almost constant within the uncertainty or may marginally be the 
largest at minimum light).  
This means that the $N$-band apparent size is largely governed by 
the opacity and the geometrical extension of the \mbox{C$_2$H$_2$}\ layers and 
the dust shell, and does not reflect the size variation of the 
underlying photosphere.  

The effects of the 
uncertainty of the effective temperature of the central star on 
the derived parameters are checked by calculating models with 
\mbox{$T_{\rm eff}$}\ = 3000 and 2600~K and included in the uncertainty 
of the derived parameters.  However, these effects have turned out to 
be relatively minor, because the \mbox{C$_2$H$_2$}\ layers, particularly the 
hot layer, are mostly optically thick, masking the underlying 
central star.  
We also examine the effects of HCN on model visibilities and spectra 
by adding the HCN contribution to the best-fit models, 
using the line list of Harris et al (\cite{harris02}) and the  
Gaussian line profile with a FWHM of 5~\mbox{km s$^{-1}$}.  
However, since the spectral feature of HCN is blended with the 
\mbox{C$_2$H$_2$}\ \mbox{$\nu_{5}$}\ band and the SiC dust feature, we cannot put a 
strong constraint on the HCN column densities, and only the 
upper limits of the HCN column densities are derived as given in 
Table~\ref{wme_param_table}.  

\begin{figure*}[!hbt]
\resizebox{\hsize}{!}{\rotatebox{0}{\includegraphics{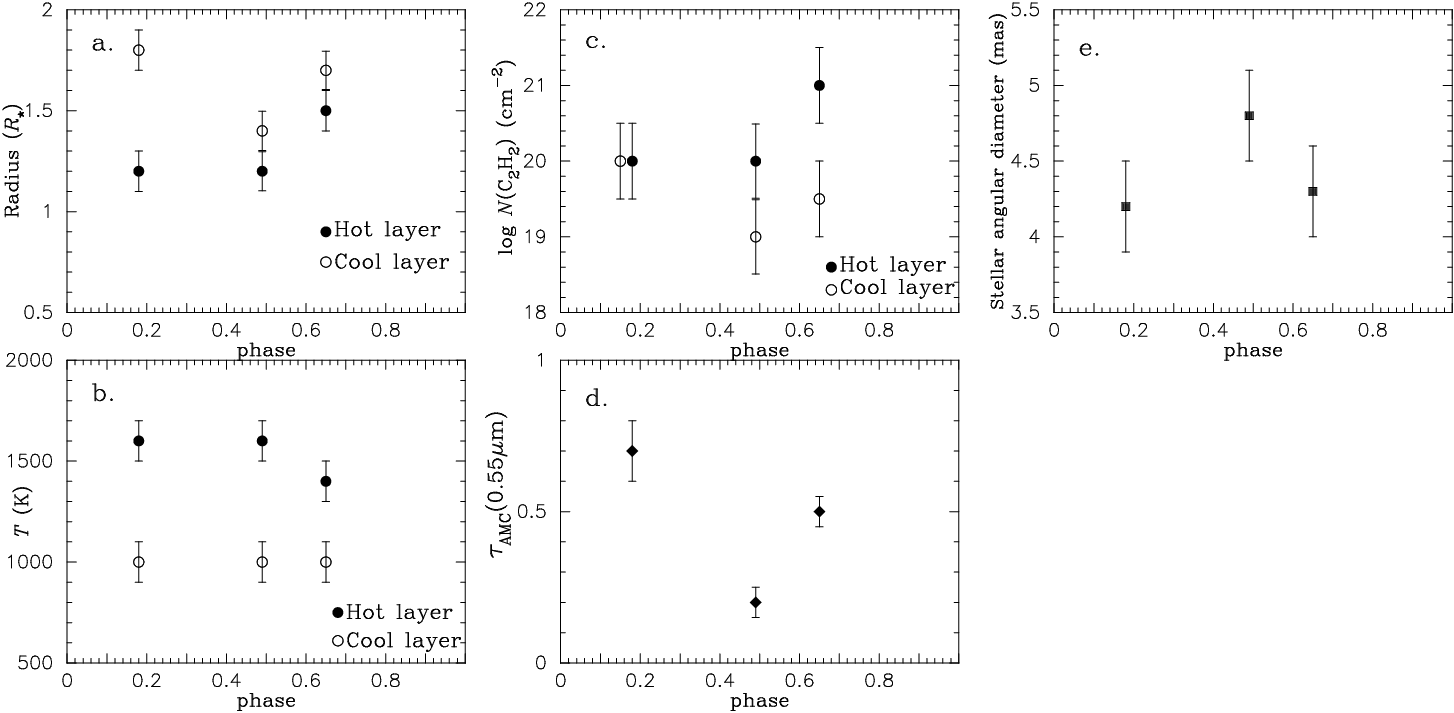}}}
\caption{Temporal variations of the radii, temperatures, and 
column densities of the hot and cool \mbox{C$_2$H$_2$}\ layers 
({\bf a}, {\bf b}, and {\bf c}), 
the optical depth of amorphous carbon dust in the dust shell 
({\bf d}), and the continuum stellar angular diameter ({\bf e}).  
}
\label{wme_param_plot}
\end{figure*}

\begin{table}
\begin{center}
\caption {Derived parameters of the \mbox{C$_2$H$_2$}\ (+HCN) layers and the dust shell 
around V~Oph.  The stellar angular diameter (2\mbox{$R_{\star}$}/$d$), where $d$ 
is the distance of V~Oph, was derived by fitting the model $N$-band 
spectra to those observed.  
}
\vspace*{-2mm}

\begin{tabular}{l c c c}\hline
         & phase 0.18      &  phase 0.49 & phase 0.65 \\ 
\hline
\mbox{$T_{\rm hot}$}\ (K)      & $1600 \pm 100$ & $1600\pm 100$  & $1400\pm 100$  \\
\mbox{$R_{\rm hot}$}\ (\mbox{$R_{\star}$}) & $1.2\pm 0.1$ & $1.2\pm 0.1$  & $1.5\pm 0.1$  \\
$\log \mbox{$N_{\rm hot}$}$ (\mbox{C$_2$H$_2$}) (\mbox{cm$^{-2}$}) & $20.0\pm 0.5$ & $20\pm 0.5$ & $21\pm 0.5$ \\
\mbox{$T_{\rm cool}$}\ (K)      & $1000 \pm 100$ & $1000\pm 100$  & $1000\pm 100$  \\
\mbox{$R_{\rm cool}$}\ (\mbox{$R_{\star}$}) & $1.8\pm 0.1$ & $1.4\pm 0.1$  & $1.7\pm 0.1$  \\
$\log \mbox{$N_{\rm cool}$}$ (\mbox{C$_2$H$_2$}) (\mbox{cm$^{-2}$}) & $20.0\pm 0.5$ & $19\pm 0.5$ & $19.5\pm 0.5$ \\
\mbox{$\tau_{V}^{\rm amc}$}\     & $0.7\pm 0.1$   &  $0.2\pm 0.05$    & $0.5\pm 0.05$  \\
$f_{\rm SiC} = \mbox{$\tau_{V}^{\rm SiC}$}/(\mbox{$\tau_{V}^{\rm amc}$} + \mbox{$\tau_{V}^{\rm SiC}$})$    
                 & $0.2\pm 0.1$   &  $0.2\pm 0.1$    & $0.2\pm 0.1$  \\
$\log \mbox{$N_{\rm hot}$}$(HCN) (\mbox{cm$^{-2}$}) & $\le 20$ & $\le 19$ & $\le 21$ \\
$\log \mbox{$N_{\rm cool}$}$(HCN) (\mbox{cm$^{-2}$}) & $\le 20$ & $\le 18$ & $\le 21$ \\
Stellar angular  & $4.2\pm 0.3$ & $4.8\pm 0.3$ & $4.3\pm 0.3$ \\
diameter (mas)   &              &              &             \\ \hline
\label{wme_param_table}
\vspace*{-7mm}

\end{tabular}
\end{center}
\end{table}

Figure~\ref{wme_param_plot} shows the derived properties of the 
\mbox{C$_2$H$_2$}\ layers and the dust shell as functions of phase, and 
a simple cartoon of the models at the three phases is shown 
in Fig.~\ref{wme_cartoon} to illustrate the size of the \mbox{C$_2$H$_2$}\ layers 
and the dust shell and their phase dependence.  
The radius of the cool \mbox{C$_2$H$_2$}\ layer 
show noticeable variations: the layer is remarkably more 
extended at post-maximum (phase 0.18) and post-minimum (phase 0.65) 
than at minimum light (phase 0.49).  
Furthermore, the column density of the cool \mbox{C$_2$H$_2$}\ layer appears 
to be the smallest at minimum light, although the uncertainties of 
the derived column densities are rather large.  The temperature of 
the cool layer is found to exhibit no temporal variation within 
the uncertainties of our modeling.  
However, it is not straightforward to interpret these 
temporal variations of the radii as the oscillating motion of the 
\mbox{C$_2$H$_2$}\ layer.  
That is, the smaller radius of the cool layer at minimum light 
(phase 0.49) does not necessarily mean that \mbox{C$_2$H$_2$}\ gas shrinks from 
phase 0.18 to 0.49 and bounces back from phase 0.49 to 0.65.  
As discussed in Sect.~\ref{sect_discuss}, it is also possible that 
\mbox{C$_2$H$_2$}\ layers may be associated with the successive passage of shock 
waves propagating outward and that the observed temporal variations 
of the radii and column densities of the \mbox{C$_2$H$_2$}\ layers are simply a 
series of ``snapshots'' of such a dynamical atmosphere.  

The variation of the \mbox{C$_2$H$_2$}\ column densities may appear to 
contradict the phase dependence of the strength of the 
8--9~\mbox{$\mu$m}\ feature mentioned in Sect.~\ref{sect_obs}: 
the absorption becomes {\em stronger} at minimum light 
(phase 0.49) than at post-maximum (phase 0.18).  
This seemingly inconsistent result can be 
explained as follows.  At maximum light, the larger geometrical 
extension of the \mbox{C$_2$H$_2$}\ layers results in the higher contribution 
of the emission from the extended part of the layers (i.e., beyond 
the central star's limb).  In other words, the absorption is 
partially, though not entirely, filled 
in by this emission component.  On the other hand, at minimum 
light, the smaller geometrical extension of the \mbox{C$_2$H$_2$}\ layers 
makes the contribution of this extended emission component 
smaller, which results in stronger absorption.  

The hot \mbox{C$_2$H$_2$}\ layer shows a temporal variation somewhat different 
from that of the cool layer.  
The hot layer becomes more extended, denser, and a little cooler 
at phase 0.65 compared to the preceding two phases.  
The dust shell also shows noticeable temporal 
variations: its optical depth is the smallest at minimum light, while 
the fraction of SiC dust does not show significant changes within 
the uncertainty of our modeling.  
This phase dependence of the optical depth of the dust shell is 
similar to that found in optically bright oxygen-rich Miras 
$o$~Cet and Z~Cyg by Suh (\cite{suh04}), who concludes that 
stronger stellar winds near maximum light (higher stellar 
luminosities) lead to higher dust production.  
We will discuss this point further in Sect.~\ref{sect_discuss}.

We note that while the inner boundary radius of the dust shell 
in the above modeling is fixed by the condensation temperature 
of amorphous carbon, the stellar pulsation 
may lead to departure from radiative equilibrium, and dust may form 
closer to the star than it would in radiative equilibrium 
(Willson \cite{willson00}).    
In order to assess this possibility, we compute dust shell models 
with an inner boundary radius equal to $\mbox{$R_{\rm cool}$}$ (i.e., no gap between 
the cool \mbox{C$_2$H$_2$}\ layer and the dust shell inner boundary) 
by artificially 
imposing an upper limit of 1600~K on the dust temperatures.  
We find out that these ad hoc 
``non-equilibrium'' models can reproduce the observed $N$-band 
visibilities and spectra at three phases if the optical depths are 
a little increased with the parameters of the \mbox{C$_2$H$_2$}\ layers kept 
the same as the above models.  For example, the \mbox{C$_2$H$_2$}\ + dust shell 
model with \mbox{$\tau_{V}^{\rm amc}$}\ = 1.0 and 
\mbox{$\tau_{V}^{\rm SiC}$}\ = 0.25 can provide a fit 
as good as the model shown in Fig.~\ref{wme_model_max}.  
This means that the inner boundary radius of the dust shell cannot be 
well constrained, because the visibility component of the dust shell, 
which appears as a steep drop at the lowest spatial frequencies, 
is not sampled by our MIDI observations using rather long baseline 
lengths.  In fact, it would also be possible that dust coexists with 
the \mbox{C$_2$H$_2$}\ gas.  In this case, the innermost part of the dust shell 
would be at least partially masked by the optically thick \mbox{C$_2$H$_2$}\ gas, 
which would mimic the dust shell with its inner boundary equal to 
\mbox{$R_{\rm cool}$}.  It is also worth noting that the non-LTE grain formation 
calculation of 
Kozasa et al. (\cite{kozasa96}) shows that SiC grains form first 
and then amorphous carbon mantles form around the SiC cores at 
low mass loss rates, which is not considered in our present model.  
These dust shell properties could be better constrained by observations 
with much shorter baselines ($\la$20~m).  

\begin{figure*}[!hbt]
\resizebox{\hsize}{!}{\rotatebox{0}{\includegraphics{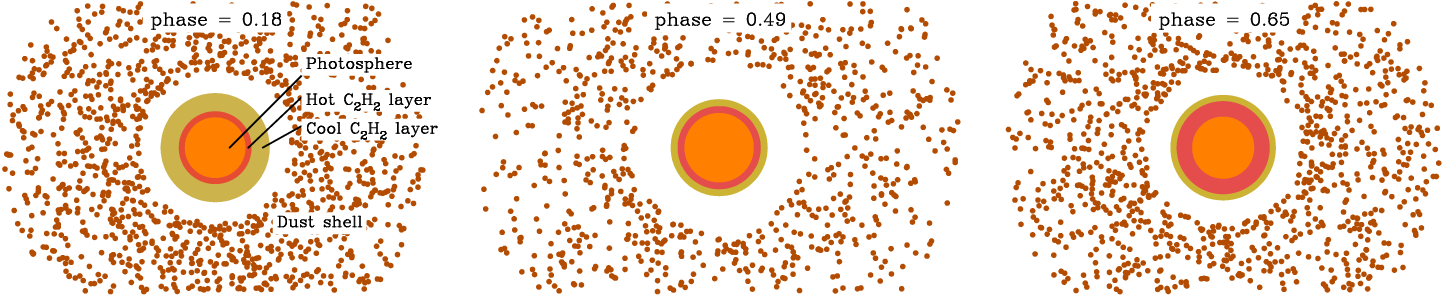}}}
\caption{A simple cartoon illustrating the best-fit models at 
phases 0.18, 0.49, and 0.65.  The sizes of the components at 
three phases {\em are to scale} except for the outer radius of 
the dust shell.  
}
\label{wme_cartoon}
\end{figure*}

There are still discrepancies between the best-fit models and the observed 
data.  At phase 0.18 and 0.65, the predicted 
visibilities (as a function spatial frequency) are still somewhat 
too flat 
compared to the MIDI visibilities measured at two baseline lengths.  
This means that the \mbox{C$_2$H$_2$}\ layers may be larger than in these models.  
In the framework 
of our two-layer modeling, however, adoption of larger radii for the 
hot and/or cool \mbox{C$_2$H$_2$}\ layers turns the absorption feature between 
8 and 9~\mbox{$\mu$m}\ to emission.  
Moreover, while we assume for simplicity that the hot \mbox{C$_2$H$_2$}\ layer 
extends down to the photosphere and the hot and cool \mbox{C$_2$H$_2$}\ layers 
are in contact, it is possible that these \mbox{C$_2$H$_2$}\ layers are detached 
and geometrically thin.  This may also explain the 
discrepancy between the models and the observed data.  
In any case, our two-layer model with 
constant temperatures and densities is a rather simplistic representation 
of real density and temperature distributions, and more sophisticated 
models of the \mbox{C$_2$H$_2$}\ layers constrained by further complimentary 
observational data may reproduce the MIDI data better.  
For example, mid-infrared high-resolution spectra of the \mbox{C$_2$H$_2$}\ 
\mbox{$\nu_{4}$+$\nu_{5}$}\ and \mbox{$\nu_{5}$}\ bands would shed 
more light on the physical 
properties of the \mbox{C$_2$H$_2$}\ layers, as 11~\mbox{$\mu$m}\ high-resolution 
spectra have proven to be strong observational constraints 
in the case of oxygen-rich Mira variables 
(e.g., Ohnaka \cite{ohnaka04a}; Ohnaka \cite{ohnaka06b}).  
Also, although we assume LTE in the calculation of the molecular opacities, 
it is necessary to examine non-LTE effects when an extensive 
high-temperature line list of \mbox{C$_2$H$_2$}\ becomes available.

\section{Discussion}
\label{sect_discuss}

Yamamura et al. (\cite{yamamura97}) analyzed the mid-infrared spectra 
of several carbon stars obtained with ISO/SWS and derived the \mbox{C$_2$H$_2$}\ 
column densities and temperatures from the 13.7~\mbox{$\mu$m}\ feature.  
The \mbox{C$_2$H$_2$}\ column densities of $10^{18\pm 0.5}$~\mbox{cm$^{-2}$}\ are somewhat 
lower than those we derived for the cool \mbox{C$_2$H$_2$}\ layer, while the 
temperatures of $1200\pm 300$~K that they derived are in fair agreement 
with those of the cool layer of our models.  It should be noted that 
they measured the strength of the 13.7~\mbox{$\mu$m}\ feature with respect to 
the local continuum levels, which are affected by the broad (hot) 
\mbox{C$_2$H$_2$}\ absorption component as well as dust emission.  Therefore, the
\mbox{C$_2$H$_2$}\ column density and temperature derived by 
Yamamura et al. (\cite{yamamura97}) may refer to the cooler, outermost 
layers.  

Recently, Nowotny et al. (\cite{nowotny05a}; \cite{nowotny05b}) have 
presented a self-consistent dynamical model including dust formation 
processes for the carbon-rich Mira S~Cep 
($\dot{M} \! \approx \! 10^{-6}$~\mbox{$M_{\sun}$~yr$^{-1}$}).  They demonstrate that 
their model can qualitatively reproduce the observed temporal variations 
of the radial velocity and line profiles of infrared molecular lines 
such as CO and CN.  The density distribution of their model shows 
a number of step-like features resulting from the formation of shock 
fronts propagating outward.  It is plausible that 
polyatomic molecules such as \mbox{C$_2$H$_2$}\ and HCN form behind shock fronts 
--- possibly at the innermost ones where density is high --- 
just as \mbox{H$_2$O}\ shells form in oxygen-rich Miras (Tej et al. \cite{tej03}).  
Therefore, it is well possible that the observed temporal variation of 
the radii of the \mbox{C$_2$H$_2$}\ layers results from changes of the positions of 
\mbox{C$_2$H$_2$}\ formation behind shock fronts.  This means that the cool layer 
extending to $\sim$1.8~\mbox{$R_{\star}$}\ at phase 0.18 moves farther outward and 
may no longer contribute to the $N$-band flux at phase 0.49.  And the 
small radius of the cool layer obtained at phase 0.49 suggests that 
a new \mbox{C$_2$H$_2$}\ formation front may just be emerging at a smaller radius 
and that this front may build up and move outward from phase 0.49 to
0.65.  

The phase dependence of the grain optical depths we derived 
for V~Oph can also be interpreted in a similar manner.   
The model presented by Nowotny et al. (\cite{nowotny05a}; 
\cite{nowotny05b}) show that dust is produced in discrete shells 
behind shock fronts where the density is sufficiently high and 
the temperature is below condensation temperature.  
If such a hot dust layer is responsible for the 
higher grain optical depths we derived for phase 0.18 (post-maximum 
light), the dust layer is expected to expand from phase 0.18 to 0.49 
(minimum light) and become diluted.  At minimum light, a new (inner) 
dust layer can be forming, but still with a low degree of 
condensation, which 
can qualitatively explain the smallest grain optical depths 
at minimum light.  When this new dust layer has fully grown, 
it can provide higher grain optical depths at post-minimum phases.  
However, as explained above, 
our present MIDI data cannot constrain the inner boundary 
radius of the dust shell strongly enough to detect the apparent 
motion of dust layers expected from this scenario.   
We also note that the mass loss rate of V~Oph is of the order of 
$10^{-8}$~\mbox{$M_{\sun}$~yr$^{-1}$}, which is by two orders of magnitude lower than 
that of S~Cep.  It is necessary to compare with 
dynamical models computed with parameters corresponding to V~Oph 
for better understanding physical processes responsible for 
molecule and dust formation close to the star: whether the extended 
\mbox{C$_2$H$_2$}\ layers can be explained by dust-driven winds triggered by 
large-amplitude stellar pulsation alone or some other mechanism is 
operating such as Alfv\'{e}n wave-driven winds (e.g., 
Vidotto \& Janteco-Pereira \cite{vidotto06}; Suzuki \cite{suzuki06}). 

The result that our two-layer models can reproduce the observed 
\mbox{C$_2$H$_2$}\ absorption feature between 8 and 9~\mbox{$\mu$m}\ implies that the 
emission from these dense, geometrically extended \mbox{C$_2$H$_2$}\ layers may be 
responsible for the observed weakness of the \mbox{C$_2$H$_2$}\ and HCN absorption 
features at 7 and 14~\mbox{$\mu$m}.  While the dust thermal emission also 
contributes to rendering these molecular features weak, the low mass loss 
rate of V~Oph suggests that the optically thick emission from the \mbox{C$_2$H$_2$}\ 
layers plays a major role at least in optically bright Mira-type 
carbon stars. 
In order to confirm this, it would be crucial to compare model 
predictions with observed strengths of the \mbox{C$_2$H$_2$}\ and HCN features and 
their temporal variations at 3~\mbox{$\mu$m}\ (accessible from the ground) 
as well as 7 and 14~\mbox{$\mu$m}\ (with Spitzer and/or AKARI).

\section{Concluding remarks}

We have carried out multi-epoch $N$-band spectro-interferometric 
observations of the carbon-rich Mira V~Oph and spatially resolved 
the object between 8 and 13~\mbox{$\mu$m}.  The observed $N$-band 
uniform-disk diameters are by a factor of 2--3 larger than the 
estimated stellar continuum size, as has already been established for 
oxygen-rich Miras.  The visibilities obtained 
at phases 0.18, 0.49, and 0.65 reveal that the object appears 
smaller at phase 0.49 (minimum light) than at phases 
0.18 and 0.65.  
Our simple modeling shows that 
the observed $N$-band visibilities and spectra can be reproduced 
by optically thick \mbox{C$_2$H$_2$}\ layers and a dust shell consisting 
of amorphous carbon and SiC.  
The cool \mbox{C$_2$H$_2$}\ layer is more extended and denser 
at phases 0.18 and 0.65 than at 0.49, while the hot \mbox{C$_2$H$_2$}\ layer 
becomes more extended and denser at phase 0.65 with no noticeable 
change between phase 0.18 and 0.49.  
This modeling also implies that the weakness of the 
\mbox{C$_2$H$_2$}\ and HCN absorption features at 7 and 14~\mbox{$\mu$m}\ observed 
in carbon-rich Miras can be 
explained by emission from these extended, dense \mbox{C$_2$H$_2$}\ 
(and HCN) layers, together with dust emission. 
Since our modeling is based on the \mbox{C$_2$H$_2$}\ opacity computed
with the RRHO band-model approximation, it is essential to examine 
the present results with an extensive high-temperature line list 
of \mbox{C$_2$H$_2$}\ in future.  However, while the \mbox{C$_2$H$_2$}\ column densities can 
be directly affected by the uncertainty of the \mbox{C$_2$H$_2$}\ opacity, the 
trend of the phase-dependent variation of the \mbox{C$_2$H$_2$}\ column densities 
found in the present work is less likely to be largely affected.  

In order to better understand the phase and cycle dependence of the 
physical properties of the outer atmosphere and the dust shell, it is 
indispensable to carry out near-infrared interferometric observations 
simultaneously with MIDI observations.  
Such coordinated observations are in fact feasible with VLTI, 
using MIDI and the near-infrared beam combiner, AMBER, which is 
capable of measuring visibilities in the $J$, $H$, and $K$ bands 
with three different spectral resolutions of 75, 1500, and 
12000.  With the highest spectral resolution, it is possible to measure 
the stellar continuum 
diameter, although not entirely free from molecular line 
contamination, and to obtain a more comprehensive picture of how the 
outer atmosphere and the dust shell respond to stellar pulsation.

\begin{acknowledgement}
We thank the ESO VLTI team on Paranal and in Garching and the MIDI 
team for carrying out the observations and making the data reduction 
software publicly available.  
We are also indebted to the referee, B.~Mennesson, for his 
constructive comments.  
We acknowledge with thanks the variable star observations from the 
AAVSO International Database contributed by observers worldwide and 
used in this research.  
\end{acknowledgement}


\begin{thebibliography}{}

\bibitem[1998]{aoki98}
Aoki, W., Tsuji, T., \& Ohnaka, K.\ 1998, A\&A, 340, 222

\bibitem[1999]{aoki99}
Aoki, W., Tsuji, T., \& Ohnaka, K.\ 1999, A\&A, 350, 945

\bibitem[2001]{bergeat01}
Bergeat, J., Knapik, A., \& Rutily, B.\ 2001, A\&A, 369, 178

\bibitem[2005a]{chesneau05a}
Chesneau, O., Meilland, A., Rivinius, T., et al.\ 2005a, A\&A, 435, 275

\bibitem[2005b]{chesneau05b}
Chesneau, O., Verhoelst, T., Lopez, B., et al.\ 2005b, A\&A, 435, 563

\bibitem[1999]{cohen99}
Cohen, M., Walker, R. G., Carter, B., et al.\ 1999, AJ, 117, 1864

\bibitem[2005]{fedele05}
Fedele, D., Wittkowski, M., Parasce, F., et al.\ 2005, A\&A, 431, 1019

\bibitem[2004]{gautchy-loidl04}
Gautchy-Loidl, R., H\"ofner, S., J\o rgensen, U. G., \& 
Hron, J.\ 2004, A\&A, 422, 289

\bibitem[2006]{groenewegen06}
Groenewegen, M. A. T.\ 2006, A\&A, 448, 181

\bibitem[1998]{groenewegen98}
Groenewegen, M. A. T., Whitelock, P. A., Smith, C. H., \& 
Kerschbaum, F.\ 1998, MNRAS, 293, 18

\bibitem[2002]{harris02}
Harris, G., Polyansky, O. L., \& Tennyson, J.\ 2002, ApJ, 578, 657

\bibitem[1979]{hinkle79}
Hinkle, K. H., \& Barnes, T. G.\ 1979, ApJ, 227, 923

\bibitem[1998]{hron98}
Hron, J., Loidl, R., H\"ofner, S., et al.\ 1998, A\&A, 335, L69

\bibitem[2004]{ireland04}
Ireland, M., Tuthill, P., Robertson, G., et al.\ 2004, In: 
Variable Stars in the Local Group, eds. D.~W.~Kurtz \& K.~Pollard, 
ASP. Conf. Series, Vol. 317, 327

\bibitem[2004]{jaffe04}
Jaffe, W.\ 2004, SPIE Proc., 5491, 715

\bibitem[2000]{jorgensen00}
J\o rgensen, U. G., Hron, J., \& Loidl, R.\ 2000, A\&A, 356, 253

\bibitem[1991]{kabbadj91}
Kabbadj, Y., Herman, M., Di Lonardo, G., \& Fusina, L.\ 1991,
J.~Mol.~Sp., 150, 535

\bibitem[2006]{kervella06}
Kervella, P., M\'{e}rand, A., Perrin, G., \& Coud\'{e} du Foresto, V.\ 
2006, A\&A, 448, 623

\bibitem[1996]{kozasa96}
Kozasa, T., Dorschner, J., Henning, Th., \& Stognienko, R.\ 1996, A\&A, 
307, 551

\bibitem[1968]{landolt68}
Landolt, A. U.\ 1968, PASP, 80, 680

\bibitem[1973]{landolt73}
Landolt, A. U.\ 1973, PASP, 85, 625

\bibitem[2004]{leinert04}
Leinert, Ch., van Boekel, R., Waters, L. B. F. M., et al.\ 2004, 
A\&A, 423, 537 

\bibitem[2002]{mennesson02}
Mennesson, B., Perrin, G., Chagnon, G., et al.\  2002, ApJ, 579, 446

\bibitem[1981]{noguchi81}
Noguchi, K., Kawara, K., Kobayashi, Y., Okuda, H., \& Sato, S.\ 1981, 
PASJ, 33, 373

\bibitem[2005a]{nowotny05a}
Nowotny, W., Aringer, B., H\"{o}fner S., Gautschy-Loidl, R., \& 
Windsteig, W.\ 2005a, A\&A, 437, 273

\bibitem[2005b]{nowotny05b}
Nowotny, W., Lebzelter, T., Hron, J., \& H\"{o}fner, S.\ 2005b, A\&A, 
437, 285

\bibitem[2004b]{ohnaka04b}
Ohnaka, K.\ 2004b, A\&A, 421, 1149

\bibitem[2004a]{ohnaka04a}
Ohnaka, K.\ 2004a, A\&A, 424, 1011

\bibitem[1996]{ohnaka96}
Ohnaka, K., \& Tsuji, T.\ 1996, A\&A, 310, 933

\bibitem[2005]{ohnaka05}
Ohnaka, K., Bergeat, J., Driebe, T., et al.\ 2005, A\&A, 429, 1057

\bibitem[2006a]{ohnaka06a}
Ohnaka, K., Driebe, T., Hofmann, K.-H., et al.\ 2006a, A\&A, 445, 1015

\bibitem[2006b]{ohnaka06b}
Ohnaka, K., Scholz, M., \& Wood, P. R.\ 2006b, A\&A, 446, 1119

\bibitem[1988]{pegourie88}
P\'{e}gouri\'{e}, B.\ 1988, A\&A, 194, 335

\bibitem[2004]{perrin04}
Perrin, G., Ridgway, S. T., Mennesson, B., et al.\ 2004, A\&A, 426, 279

\bibitem[2006]{poncelet06}
Poncelet, A., Perrin, G., \& Sol, H.\ 2006, A\&A, 450, 483

\bibitem[2003]{przygodda03}
Przygodda, F., Chesneau, O., Graser, U., Leinert, Ch., \& 
Morel, S.\ 2003, Ap\&SS, 286, 85

\bibitem[2006]{quanz06}
Quanz, S. P., Henning, Th., Bouwman, J., Ratzka, Th., \& Leinert, Ch.\ 
2006, ApJ, 648, 472

\bibitem[1991]{rouleau91}
Rouleau, F., \& Martin, P. G.\ 1991, ApJ, 377, 526

\bibitem[2004]{samus04}
Samus, N. N., Durlevich, O. V, et al.\ 2004, 
Combined General Catalogue of Variable Stars

\bibitem[1979]{savage79}
Savage, B. D., \& Mathis, J.S.\ 1979, ARA\&A, 17, 73

\bibitem[2004]{suh04}
Suh, K.-W.\ 2004, ApJ, 615, 485

\bibitem[2006]{suzuki06}
Suzuki, T. K.\ 2006, submitted to ApJ, astro-ph/0608195

\bibitem[2003]{tej03}
Tej, A., Lan\c{c}on, A., \& Scholz, M.\ 2003, A\&A, 401, 347

\bibitem[2002]{thompson02}
Thompson, R. R., Creech-Eakman, M. J., \& van Belle, G. T.\ 2002, 
ApJ, 577, 447

\bibitem[1984]{tsuji84}
Tsuji, T.\ 1984, A\&A, 134, 24

\bibitem[1997]{tsuji97}
Tsuji, T., Ohnaka K., Aoki, W., \& Yamamura, I.\ 1997, A\&A, 320, L1

\bibitem[1997]{vanbelle97}
van Belle, G. T., Dyck, H. M., Thompson, R. R., Benson, J. A., \& 
Kannappan, S. J.\ 1997, AJ, 114, 2150

\bibitem[2006]{vidotto06}
Vidotto, A. A., \& Janteco-Pereira, V.\ 2006, ApJ, 639, 416

\bibitem[1979]{walker79}
Walker, A. R.\ 1979, SAAO Circ. 1, 112

\bibitem[2000]{whitelock00}
Whitelock, P., Marang, F., \& Feast, M.\ 2000, MNRAS, 319, 728

\bibitem[2000]{willson00}
Willson, L. A.\ 2000, ARA\&A, 38, 573

\bibitem[2004]{woodruff04}
Woodruff, C., Eberhardt, M., Driebe, T., et al.\ 2004, A\&A, 421, 703

\bibitem[1997]{yamamura97}
Yamamura, I., de Jong, T., Justtanont, K., Cami, J., \& 
Waters, L. B. F. M.\ 1997, Ap\&SS, 255, 351

\bibitem[1999]{yamamura99}
Yamamura, I., de Jong, T., \& Cami, J.\ 1999, A\&A, 348, L55

\end{thebibliography}
\end{document}